\documentclass[10pt,conference]{IEEEtran}
\usepackage{mathptmx}
\usepackage{graphicx}

\usepackage{amsmath}
\usepackage{algorithm}
\usepackage[noend]{algpseudocode}
\usepackage{tabulary}
\usepackage{tikz}
\usetikzlibrary{calc}
\usepackage{pgfplots}
\usepackage{titletoc}
\usepackage{pgfgantt}
\usepackage{gillius2}
\usepackage{booktabs}
\usepackage{afterpage}
\usepackage{placeins}
\usepackage{url}
\usepackage{inconsolata}
\usepackage{multirow}
\usepackage[colorinlistoftodos]{todonotes}
\usepackage[framemethod=TikZ]{mdframed}
\newmdenv[hidealllines=true,backgroundcolor=black!10,skipabove=1pt,skipbelow=2pt,linewidth=0pt]{shaded}
\usepackage[numbers]{natbib}
\usetikzlibrary{decorations.pathmorphing}
\usetikzlibrary{positioning}
\usepackage[eulergreek]{sansmath}
\let\labelindent\relax
\usepackage{enumitem}
\usepackage{fontawesome}
\usepackage{fancyhdr}

\definecolor{colorSG}{RGB}{168,50,45}
\definecolor{customY}{HTML}{FBB13C}
\definecolor{customG}{HTML}{218380}
\definecolor{customT}{HTML}{73D2DE}
\definecolor{customW}{HTML}{FCFCFF}
\definecolor{customB}{HTML}{2E5EAA}
\definecolor{customP}{HTML}{5B4E77}
\definecolor{gitRedFaded}{HTML}{ffeef0}
\definecolor{gitGreenFaded}{HTML}{e6ffed}
\definecolor{gitRed}{HTML}{ffdce0}
\definecolor{gitGreen}{HTML}{cdffd8}
\definecolor{gitRedFull}{HTML}{cb2431}
\definecolor{gitGreenFull}{HTML}{2cbe4e}

\input{tools/tools_entity_relationship_diagrams.tex}

\begin{document}

\title{git2net - Mining Time-Stamped Co-Editing Networks from Large git Repositories}

\author{\IEEEauthorblockN{Christoph Gote}
\IEEEauthorblockA{\textit{Chair of Systems Design}\\
\textit{ETH Z\"urich} \\
Zurich, Switzerland \\
cgote@ethz.ch}
\and
\IEEEauthorblockN{Ingo Scholtes}
\IEEEauthorblockA{\textit{Data Analytics Group}\\
	\textit{Department of Informatics}\\
\textit{University of Zurich}\\
Zurich, Switzerland \\
scholtes@ifi.uzh.ch\vspace*{-5mm}}
\and
\IEEEauthorblockN{Frank Schweitzer}
\IEEEauthorblockA{\textit{Chair of Systems Design}\\
\textit{ETH Z\"urich}\\
Zurich, Switzerland \\
fschweitzer@ethz.ch}
}

\maketitle

\fancyhf{} 
\renewcommand{\headrulewidth}{0pt}
\cfoot{\sffamily\footnotesize\bfseries\copyright~2019 IEEE. Personal use of this material is permitted. Permission from IEEE must be obtained for all other uses, in any current\\
or future media, including reprinting/republishing this material for advertising or promotional purposes, creating new collective\\
works, for resale or redistribution to servers or lists, or reuse of any copyrighted component of this work in other works.}
\thispagestyle{fancy}


\begin{abstract}
Data from software repositories have become an important foundation for the empirical study of software engineering processes.
A recurring theme in the repository mining literature is the inference of developer networks capturing e.g. collaboration, coordination, or communication from the commit history of projects.
Most of the studied networks are based on the \emph{co-authorship} of software artefacts defined at the level of files, modules, or packages.
While this approach has led to insights into the social aspects of software development, it neglects detailed information on code changes and code ownership, e.g. which exact lines of code have been authored by which developers, that is contained in the commit log of software projects.

Addressing this issue, we introduce \texttt{git2net}, a scalable \texttt{python} software that facilitates the extraction of fine-grained \emph{co-editing networks} in large \texttt{git} repositories.
It uses text mining techniques to analyse the detailed history of textual modifications \emph{within} files. 
This information allows us to construct directed, weighted, and time-stamped networks, where a link signifies that one developer has edited a block of source code originally written by another developer.
Our tool is applied in case studies of an Open Source and a commercial software project.
We argue that it opens up a massive new source of high-resolution data on human collaboration patterns.
\end{abstract}

\section{Introduction}

Software repositories are a rich source of data facilitating empirical studies of software engineering processes.
Methods to use meta-data from these repositories have become a common theme in the repository mining literature.
Thanks to the availability of massive databases, already simple means allow to query meta-data on the commits of developers~\cite{gousios2012ghtorrent,gousios2017mining}.
Apart from the evolution of software artefacts, they also contain a wealth of fine-grained information on the human and social aspects of software development teams.
Specifically, the commit history of developers allows to construct social networks that proxy collaboration, coordination, or communication structures in software teams.
These databases have therefore facilitated data-driven studies of social systems not only in empirical software engineering, but also in areas like computational social science, social network analysis, organisational theory, or management science~\cite{carley2001computational,vonKrogh2006}.


The detailed record of file modifications contained in the commit log of, e.g. \texttt{git} repositories also enables more advanced network reconstruction techniques.
In particular, from the micro-level analysis of textual modifications between subsequent versions of code we can infer \emph{time-stamped, weighted, and directed co-editing relationships}.
Such a relationship $(A,B; t, w)$ indicates that at time $t$ developer $A$ modified $w$ characters of code originally written by another developer $B$.
Recent research has shown that such a fine-grained analysis of co-editing networks in large software projects can provide insights that go beyond more coarse-grained definitions~\cite{Joblin2015,Scholtes2016}.
However, a tool to conveniently extract such rich, time-stamped collaboration networks for the large corpus of \texttt{git} repositories available, e.g. via public platforms like \texttt{gitHub}, is currently missing.

Addressing this gap, we present such a tool that facilitates the scalable extraction of time-stamped co-editing relationships between developers in large software repositories.
The contributions of our work are as follows:
\begin{itemize}[leftmargin=*]
    \item[\raisebox{.3ex}{\tiny\faPlay}] We introduce \texttt{git2net}, a python tool that can be used to mine time-stamped co-editing relations between developers from the sequence of file modifications contained in \texttt{git} repositories.
    Building on the repository mining framework \texttt{pyDriller}~\cite{PyDriller}, \texttt{git2net} can operate both on local and remote repositories.
    Providing a command-line interface as well as an API, \texttt{git2net} can be used as stand-alone tool for standard analysis tasks as well as a framework for the implementation of advanced data mining scripts.
    Our tool is available as an Open Source project\footnote{\url{https://github.com/gotec/git2net}}.
    \item[\raisebox{.3ex}{\tiny\faPlay}] Analysing all file modifications contained in the commit log, \texttt{git2net} generates a database that captures fine-grained information on co-edited code either at the level of lines or contiguous code regions.
    Building on text mining techniques, it further analyses the overlap between co-edited code regions using (i) the Levenshtein edit distance~\cite{levenshtein1966} and (ii) a text-based entropy measure~\cite{shannon1948mathematical}.
    These measures facilitate (i) a character-based proxy estimating the effort behind code modifications, and (ii) an entropy-based correction for binary file changes that can have a considerable impact on text-based effort estimation techniques.
    \item[\raisebox{.3ex}{\tiny\faPlay}] We develop an approach to generate time-stamped collaboration networks based on multiple projections: (i) time-stamped co-editing networks, (ii) time-stamped bipartite networks linking developers to edited files, and (iii) directed acyclic graphs of code edits that allow to infer ``paths'' of consecutive edits building upon each other.
    All network projections are implemented in \texttt{git2net} and can be directly exported as \texttt{HTML} visualisations as well as formats readable by common network analysis tools.
    \item[\raisebox{.3ex}{\tiny\faPlay}] Thanks to a parallel processing model that utilises modern multi-core architectures, \texttt{git2net} supports the analysis of massive software repositories with hundreds of thousands of commits and millions of lines of code.
    A scalability analysis proves that our parallel implementation yields a linear speed-up compared to a single-threaded implementation, thus facilitating the fine-grained textual analysis even in massive projects with a long history.
    \item[\raisebox{.3ex}{\tiny\faPlay}] Utilizing \texttt{git2net} in a case study on two software projects, we show that the fine-grained textual analysis of file modifications yields considerably different network structures compared to coarse-grained methods that analyse code co-authorship at the level of files or modules. We further demonstrate how our tool can be used to breakdown developer effort into (a) the revision of code authored by the developer him or herself vs. (b) the revision of code written by other team members.
\end{itemize}

Providing a novel method to mine fine-grained collaboration networks at high temporal resolution from any \texttt{git} repository, our work opens new perspective for empirical studies of development processes. 
It further contributes a simple method to generate data on temporal social networks that are of interest for researchers in computational social science, (social) network analysis and organisational theory.

The remainder of this paper is structured as follows:
Section \ref{sec:related} provides an overview of works addressing the construction of social networks from software repository data.
Section \ref{sec:method} introduces our proposed methodology to extract time-resolved and directed links between developers who subsequently edit each others' code.
Section \ref{sec:results} presents a case study, in which we apply our tool to \texttt{git} repositories from (i) an Open Source Software project, and (ii) a commercial, closed-source project.
In section \ref{sec:conclusion} we draw conclusions from our work and highlight the next steps in our research.

\section{Related Work}
\label{sec:related}


Given the large body of work using network analysis to study software development processes, we restrict our overview to related works that address the reconstruction of social networks from software repositories.
A broader view on applications of graph-based data analysis and modelling techniques in empirical software engineering---including works on (technical) dependency networks that are outside the scope of our work---is, e.g., available in \cite{wolf2009mining,xie2009data,Cataldo2014_acs}.

A number of studies use operational data on software projects to construct graphs or networks where nodes capture developers while links capture social interactions and/or work dependencies between developers.
To this end, a first line of works has used data that directly capture communication \citep{Geipel2014}, e.g. via IRC channels~\cite{Cataldo2008}, E-Mail exchanges~\cite{bird2006mining,Wolf2009,bacchelli2011miler,hong2011understanding,Xuan2014}, mailing lists~\cite{guzzi2013communication}, or communication via issue trackers~\cite{long2007social,howison2006social,sureka2011using,Zanetti2013_icse,Scholtes2013}.

While data on direct developer communication facilitate the construction of meaningful social networks, they are often not available, e.g. due to privacy concerns.
To address such settings, researchers have developed methods to \emph{infer} or \emph{reconstruct} collaboration networks based on developer actions recorded in code repositories like \texttt{CVS}, \texttt{SVN}, or \texttt{git}.
A common approach starts from \emph{code authorship} or \emph{code ownership} networks, which map the relation between a developer and the artefacts (i.e. files, modules, binaries, etc.) that he or she contributed to~\cite{Fritz2007,Bird2011,Greiler2015,maclean2013apache}.
The resulting directed bipartite developer-artefact networks \cite{newman2018networks} can then be projected onto \emph{co-authorship networks}, where undirected links between two developers $A$ and $B$ indicate that $A$ and $B$ have modified at least one common artefact.
The authors of \cite{geipel2009software,geipel2012modularity} have studied co-change based on a large corpus of CVS repositories of Open Source Software projects.

The majority of works mining social networks from software repositories build on this general idea.
In \cite{maclean2013apache, madey2002,Meneely2008,ogawa2010software,vijayaraghavan2015quantifying} a file-based notion of co-authorship is used to construct \emph{co-commit networks}, where a link between two developers signifies that they have committed the same file at least once.
The authors of \cite{lopez2004} 
adopt a module-based definition, assuming that two developers are linked in the co-authorship network if they have contributed to at least one common module.
Taking a similar approach, Huang and Liu \citep{Huang2005} use information on modified file paths in \texttt{SourceForge} repositories to infer relations between authors editing the same part of a project.
Incorporating the time stamps of commits, Pohl and Diehl \citep{pohl2008dynamic} used a file-based co-authorship definition to construct \emph{dynamic} developer networks~ that can be analysed and visualised using methods from dynamic network analysis~\cite{Holme2015}.
The authors of \citep{Cohen2018} recently developed a similar approach to study the ecosystem of software projects on \texttt{gitHub}.
To this end, they define project-level co-commit networks, i.e. a projection of commits where two developers are linked if they committed to the same Open Source project.
Schweitzer et al. \citep{schweitzer2014oss} provided a related study, analysing ten years of data from the Open Source project hosting platform \texttt{SourceForge}.


These works have typically constructed \emph{undirected co-authorship networks} based on joint contributions to files, modules, or projects.
Such coarse-grained definitions of co-authorship networks introduce a potential issue:
They do not distinguish between (i) links between developers that are due to \emph{independent} contributions to the same artefact, and (ii) links that are due to commit sequences where one developer builds upon and/or redacts the particular lines of source code previously authored by another developer.
Networks defined based on the latter type of \emph{time-ordered co-editing} of code regions are likely associated with a stronger need for coordination and communication than the mere fact that developers edited the same file or module~\cite{cataldo2006identification}.
So far, few studies have adopted such fine-grained approaches to create developer collaboration networks. 
Notable exceptions include the function-level co-editing networks constructed by Joblin et al. \cite{Joblin2015}.
The authors further argue that, using file-based definitions of collaboration networks, network analytic methods fail to identify meaningful communities.
The authors of \cite{Scholtes2016} constructed line-based co-editing networks, showing that such an analysis (i) yields insights into the coordination structures of software teams, and (ii) provides new ways to test long-standing hypotheses about cooperative work from social psychology.

While such a fine-grained analysis of the co-editing behaviour of developers has its advantages, it also introduces challenges that have so far limited its adoption.
First and foremost, it requires a detailed analysis of file modifications and makes it necessary to identify the original author for every modified line of code affected in each commit.
Requiring a potentially large number of \texttt{git} operations for every commit being analysed, such an analysis is both complicated to implement as well as time-consuming to perform.
Compared to other approaches, which often merely require a suitable query in structured databases like \texttt{ghTorrent}~\cite{gousios2012ghtorrent,gousios2017mining}, a tool that facilitates this task for very large repositories is still missing.

Closing this gap, our work introduces a practical and scalable solution for the construction of fine-grained and time-stamped co-editing networks from \texttt{git} repositories.
Our work extends the state-of-the-art and facilitates analyses of developer collaboration and coordination in software projects.
Providing a new method to construct large, dynamic networks at high temporal resolution we further expect our work to be of interest for the community of researchers developing methods to analyse dynamic (social) networks~\cite{Holme2015, berger2006framework,carley2012dynamic}.

\section{Mining Co-Editing Relations from git Repositories}
\label{sec:method}


\subsection{From Commit Logs to Co-Edits}

We first outline our proposed method to extract co-editing relationships from \texttt{git} commits.
An overview of the mining procedure, which we will explain in the following, is presented in Algorithm \ref{alg:mining_procedure}.

\texttt{git} projects generally consist of multiple files that can be edited by a large number of developers.
Sets of changes made by a developer to potentially multiple files are recorded as commits, where each commit is identified by a unique hash. 
Building on the package \texttt{pydriller} \cite{PyDriller}, we first extract the history of all commits in a repository and record the meta-data (author, time of commit, branch, etc.) for each commit.
As the person committing the changes is not necessarily the author of these changes (a different developer can commit code on behalf of the original author), both the committer and author of the changes are considered.
Subsequently we analyse the changes made with the commit.

As each commit can contain modifications of multiple files, we analyse each file modification individually to associate every changed text region with its original author.
In a first step, select the modifications relevant for the current analysis.
To this end, we have implemented a filter allowing to exclude specific files, file types as well as entire directories or subdirectories from the analysis.
For all selected modifications, the associated \texttt{diff} is analysed, determining which lines have been added or deleted.
In addition, we identify the original author of every edited line of code by executing \texttt{git blame} on the version of the analysed file before the current commit.
By matching the author $A$ of a modification contained in the current commit with time stamp $t$ to all original authors $B_i$ of an edited line $i$, we obtain time-stamped and directed co-editing relations $(A, B_i, t)$.

\begin{algorithm}[t!]\footnotesize
\caption{Simplified mining procedure of \texttt{git2net}}\label{alg:mining_procedure}
\begin{algorithmic}[1]
\Procedure{mine\_git\_repo}{git\_repo, output\_db}
\For {all commits in git\_repo}
	\State commit\_info $\gets$ parsed commit data
	\For {all modified files in commit}
		\State deleted\_lines, added\_lines $\gets$ parse diff of modification
		\State blame\_info $\gets$ git blame on file in parent commit
		\For {each line deleted lines}
			\State current\_author $\gets$ modifying author from commit\_info
			\State previous\_author $\gets$ original author from blame\_info
			\State coedits\_info $\gets$ authors and metadata on changes
		\EndFor
	\EndFor
	\State output\_db $\gets$ commit\_info, coedits\_info
\EndFor
\EndProcedure
\end{algorithmic}
\end{algorithm}

For each extracted relation, we record hashes of the original and modifying commit as well as meta-data capturing the location (file name, line number) of the associated co-edit.
Naturally, such co-edits can be linked to vastly different development effort, ranging from a change of whitespaces to the complete rewriting of code.
To capture to what extent developers edit each others' code, we use a text mining approach to address these differences.
We specifically use the Levenshtein edit distance \cite{levenshtein1966}, which can be thought of as the minimum number of keystrokes required to transform the prior source code version into the version after the edit. 
This measure proxies the development effort associated with an edit, where single character changes, line deletions, or the commenting/uncommenting of lines are associated with a minimum effort while the writing of a new line of code is associated with maximum effort.
This approach allows us to construct time-stamped and \emph{weighted} co-edit relations $(A,B;t,w)$, where the weight $w$ captures the Levenshtein distance of the associated edit.


An issue that we have encountered during the testing of our method in real-world repositories is associated with the embedding of text-encoded binary objects in source code, e.g. due to the inclusion of \texttt{base64}-encoded images in \texttt{HTML} or \texttt{JavaScript}.
Notably, the modification of a single pixel in a text-encoded image, can result in a completely different text encoding.
Considering our approach to associate the weight of a co-edit relation with the Levensthein edit distance this can considerably distort our analysis, potentially leading to the issue that binary file modifications dominate the recorded weights.
We take an information-theoretic approach to enable the detection (and potential exclusion) of such modifications.
In particular, we compute the entropy $S$ of code before and after the change, defined as: 
\begin{align}
S = -\sum_k \textbf{p}_k \log_2(\textbf{p}_k)
\end{align}
This computation is based on the \texttt{utf-8} encoding space with 256 possible symbols. 
Entries of the vector $\textbf{p}$ represent a symbol's normalised frequency in a given string. Given this definition, the entropy $S$ can take values between 0 and 8 bits.
Some examples for this measure are given in Figure \ref{fig:entropy_example}. 
The resulting distribution of entropy for all co-edits can be used for a Bayesian classification distinguishing, e.g. binary encoded images or hashes from natural language or source code changes.

\begin{figure}[t!]
    \centering
    \sffamily
	\begin{tabular}{clc}
		& \textbf{code} & \textbf{entropy}\\
		\cmidrule{2-3}
		\textbf{a} & \texttt{for x in 'hello world': print(x)} & 3.94 \\
		\textbf{b} & \texttt{for c in 'hello world': print(c)} & 3.94 \\
		\textbf{c} & \texttt{d = \{x[0]:x[1] for x in df['d']\}} & 3.80\\
		\textbf{d} & \texttt{Uatsffm+BC+s7kWKqVpMlrMEWk7nTfK1} & 4.41 \\
		&&\\ 
	\end{tabular}
	\caption{Entropy of equal length strings based on discrete \texttt{utf-8} (256 possible symbols) probability space. The entropy can take values between 0 and 8 bits. The entropy of \texttt{base64} encoded image (\textbf{\sffamily d}) is considerably higher than of typical lines of (\texttt{python}) code (\textbf{\sffamily a}--\textbf{\sffamily c}). In practice the effect is amplified as strings of binary encoded images are longer. Small changes within a line have a small or no effect on entropy as can be seen in the entropy difference between \textbf{\sffamily a} and \textbf{\sffamily b}.}\label{fig:entropy_example}
\end{figure}

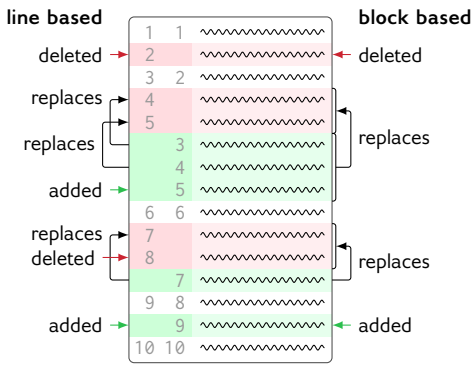
\begin{figure}[b!]
    \centering
    \begin{tikzpicture}\ttfamily\footnotesize
    	\coordinate (pre1) at (0,0);
    	\coordinate[node distance=3mm, below=of pre1.east, anchor=east] (pre2);
    	\coordinate[node distance=3mm, below=of pre2.east, anchor=east] (pre3);
    	\coordinate[node distance=3mm, below=of pre3.east, anchor=east] (pre4);
    	\coordinate[node distance=3mm, below=of pre4.east, anchor=east] (pre5);
    	\coordinate[node distance=3mm, below=of pre5.east, anchor=east] (pre6);
    	\coordinate[node distance=3mm, below=of pre6.east, anchor=east] (pre7);
    	\coordinate[node distance=3mm, below=of pre7.east, anchor=east] (pre8);
    	\coordinate[node distance=3mm, below=of pre8.east, anchor=east] (pre9);
    	\coordinate[node distance=3mm, below=of pre9.east, anchor=east] (pre10);
    	\coordinate[node distance=3mm, below=of pre10.east, anchor=east] (pre11);
    	\coordinate[node distance=3mm, below=of pre11.east, anchor=east] (pre12);
    	\coordinate[node distance=3mm, below=of pre12.east, anchor=east] (pre13);
    	\coordinate[node distance=3mm, below=of pre13.east, anchor=east] (pre14);
    	\coordinate[node distance=3mm, below=of pre14.east, anchor=east] (pre15);
    	
    	\coordinate[node distance=4mm, right=of pre1.east, anchor=west] (post1);
    	\coordinate[node distance=3mm, below=of post1.east, anchor=east] (post2);
    	\coordinate[node distance=3mm, below=of post2.east, anchor=east] (post3);
    	\coordinate[node distance=3mm, below=of post3.east, anchor=east] (post4);
    	\coordinate[node distance=3mm, below=of post4.east, anchor=east] (post5);
    	\coordinate[node distance=3mm, below=of post5.east, anchor=east] (post6);
    	\coordinate[node distance=3mm, below=of post6.east, anchor=east] (post7);
    	\coordinate[node distance=3mm, below=of post7.east, anchor=east] (post8);
    	\coordinate[node distance=3mm, below=of post8.east, anchor=east] (post9);
    	\coordinate[node distance=3mm, below=of post9.east, anchor=east] (post10);
    	\coordinate[node distance=3mm, below=of post10.east, anchor=east] (post11);
    	\coordinate[node distance=3mm, below=of post11.east, anchor=east] (post12);
    	\coordinate[node distance=3mm, below=of post12.east, anchor=east] (post13);
    	\coordinate[node distance=3mm, below=of post13.east, anchor=east] (post14);
    	\coordinate[node distance=3mm, below=of post14.east, anchor=east] (post15);
    	
    	\draw[draw=none, fill=gitRed] ($(pre2.east)+(-4.5mm,1.5mm)$) |- ($(post2.east)+(1mm,-1.5mm)$) |- cycle;
    	\draw[draw=none, fill=gitRedFaded] ($(post2.east)+(0,1.5mm)$) |- ($(post2.east)+(1.85,-1.5mm)$) |- cycle;
    	
    	\draw[draw=none, fill=gitRed] ($(pre4.east)+(-4.5mm,1.5mm)$) |- ($(post4.east)+(1mm,-1.5mm)$) |- cycle;
    	\draw[draw=none, fill=gitRedFaded] ($(post4.east)+(0,1.5mm)$) |- ($(post4.east)+(1.85,-1.5mm)$) |- cycle;
    	\draw[draw=none, fill=gitRed] ($(pre5.east)+(-4.5mm,1.5mm)$) |- ($(post5.east)+(1mm,-1.5mm)$) |- cycle;
    	\draw[draw=none, fill=gitRedFaded] ($(post5.east)+(0,1.5mm)$) |- ($(post5.east)+(1.85,-1.5mm)$) |- cycle;
    	\draw[draw=none, fill=gitGreen] ($(pre6.east)+(-4.5mm,1.5mm)$) |- ($(post6.east)+(1mm,-1.5mm)$) |- cycle;
    	\draw[draw=none, fill=gitGreenFaded] ($(post6.east)+(0,1.5mm)$) |- ($(post6.east)+(1.85,-1.5mm)$) |- cycle;
    	\draw[draw=none, fill=gitGreen] ($(pre7.east)+(-4.5mm,1.5mm)$) |- ($(post7.east)+(1mm,-1.5mm)$) |- cycle;
    	\draw[draw=none, fill=gitGreenFaded] ($(post7.east)+(0,1.5mm)$) |- ($(post7.east)+(1.85,-1.5mm)$) |- cycle;
    	\draw[draw=none, fill=gitGreen] ($(pre8.east)+(-4.5mm,1.5mm)$) |- ($(post8.east)+(1mm,-1.5mm)$) |- cycle;
    	\draw[draw=none, fill=gitGreenFaded] ($(post8.east)+(0,1.5mm)$) |- ($(post8.east)+(1.85,-1.5mm)$) |- cycle;
    	
    	\draw[draw=none, fill=gitRed] ($(pre10.east)+(-4.5mm,1.5mm)$) |- ($(post10.east)+(1mm,-1.5mm)$) |- cycle;
    	\draw[draw=none, fill=gitRedFaded] ($(post10.east)+(0,1.5mm)$) |- ($(post10.east)+(1.85,-1.5mm)$) |- cycle;
    	\draw[draw=none, fill=gitRed] ($(pre11.east)+(-4.5mm,1.5mm)$) |- ($(post11.east)+(1mm,-1.5mm)$) |- cycle;
    	\draw[draw=none, fill=gitRedFaded] ($(post11.east)+(0,1.5mm)$) |- ($(post11.east)+(1.85,-1.5mm)$) |- cycle;
    	\draw[draw=none, fill=gitGreen] ($(pre12.east)+(-4.5mm,1.5mm)$) |- ($(post12.east)+(1mm,-1.5mm)$) |- cycle;
    	\draw[draw=none, fill=gitGreenFaded] ($(post12.east)+(0,1.5mm)$) |- ($(post12.east)+(1.85,-1.5mm)$) |- cycle;
    	
    	\draw[draw=none, fill=gitGreen] ($(pre14.east)+(-4.5mm,1.5mm)$) |- ($(post14.east)+(1mm,-1.5mm)$) |- cycle;
    	\draw[draw=none, fill=gitGreenFaded] ($(post14.east)+(0,1.5mm)$) |- ($(post14.east)+(1.85,-1.5mm)$) |- cycle;
    	
    	\draw[black!70, rounded corners=2] ($(pre1)+(-4.5mm, 2mm)$) |- ($(post15)+(1.85,-2mm)$) -- ($(post1)+(1.85,2mm)$) -- cycle;
    	
    	\path[decoration={snake,segment length=3,amplitude=.9}] ($(post1.east)+(.1,0)$) edge[decorate] ($(post1.east)+(1.75,0)$);
    	\path[decoration={snake,segment length=3,amplitude=.9}] ($(post2.east)+(.1,0)$) edge[decorate] ($(post2.east)+(1.75,0)$);
    	\path[decoration={snake,segment length=3,amplitude=.9}] ($(post3.east)+(.1,0)$) edge[decorate] ($(post3.east)+(1.75,0)$);
    	\path[decoration={snake,segment length=3,amplitude=.9}] ($(post4.east)+(.1,0)$) edge[decorate] ($(post4.east)+(1.75,0)$);
    	\path[decoration={snake,segment length=3,amplitude=.9}] ($(post5.east)+(.1,0)$) edge[decorate] ($(post5.east)+(1.75,0)$);
    	\path[decoration={snake,segment length=3,amplitude=.9}] ($(post6.east)+(.1,0)$) edge[decorate] ($(post6.east)+(1.75,0)$);
    	\path[decoration={snake,segment length=3,amplitude=.9}] ($(post7.east)+(.1,0)$) edge[decorate] ($(post7.east)+(1.75,0)$);
    	\path[decoration={snake,segment length=3,amplitude=.9}] ($(post8.east)+(.1,0)$) edge[decorate] ($(post8.east)+(1.75,0)$);
    	\path[decoration={snake,segment length=3,amplitude=.9}] ($(post9.east)+(.1,0)$) edge[decorate] ($(post9.east)+(1.75,0)$);
    	\path[decoration={snake,segment length=3,amplitude=.9}] ($(post10.east)+(.1,0)$) edge[decorate] ($(post10.east)+(1.75,0)$);
    	\path[decoration={snake,segment length=3,amplitude=.9}] ($(post11.east)+(.1,0)$) edge[decorate] ($(post11.east)+(1.75,0)$);
    	\path[decoration={snake,segment length=3,amplitude=.9}] ($(post12.east)+(.1,0)$) edge[decorate] ($(post12.east)+(1.75,0)$);
    	\path[decoration={snake,segment length=3,amplitude=.9}] ($(post13.east)+(.1,0)$) edge[decorate] ($(post13.east)+(1.75,0)$);
    	\path[decoration={snake,segment length=3,amplitude=.9}] ($(post14.east)+(.1,0)$) edge[decorate] ($(post14.east)+(1.75,0)$);
    	\path[decoration={snake,segment length=3,amplitude=.9}] ($(post15.east)+(.1,0)$) edge[decorate] ($(post15.east)+(1.75,0)$);
    	
    	\node[anchor=east, black!40] at (pre1) {1};
    	\node[anchor=east, black!40] at (pre2) {2};
    	\node[anchor=east, black!40] at (pre3) {3};
    	\node[anchor=east, black!40] at (pre4) {4};
    	\node[anchor=east, black!40] at (pre5) {5};
    	\node[anchor=east, black!40] at (pre6) {};
    	\node[anchor=east, black!40] at (pre7) {};
    	\node[anchor=east, black!40] at (pre8) {};
    	\node[anchor=east, black!40] at (pre9) {6};
    	\node[anchor=east, black!40] at (pre10) {7};
    	\node[anchor=east, black!40] at (pre11) {8};
    	\node[anchor=east, black!40] at (pre12) {};
    	\node[anchor=east, black!40] at (pre13) {9};
    	\node[anchor=east, black!40] at (pre14) {};
    	\node[anchor=east, black!40] at (pre15) {10};
    	
    	\node[anchor=east, black!40] at (post1) {1};
    	\node[anchor=east, black!40] at (post2) {};
    	\node[anchor=east, black!40] at (post3) {2};
    	\node[anchor=east, black!40] at (post4) {};
    	\node[anchor=east, black!40] at (post5) {};
    	\node[anchor=east, black!40] at (post6) {3};
    	\node[anchor=east, black!40] at (post7) {4};
    	\node[anchor=east, black!40] at (post8) {5};
    	\node[anchor=east, black!40] at (post9) {6};
    	\node[anchor=east, black!40] at (post10) {};
    	\node[anchor=east, black!40] at (post11) {};
    	\node[anchor=east, black!40] at (post12) {7};
    	\node[anchor=east, black!40] at (post13) {8};
    	\node[anchor=east, black!40] at (post14) {9};
    	\node[anchor=east, black!40] at (post15) {10};
    	
    	\sffamily\footnotesize
    	\draw[-latex, draw=gitRedFull] ($(pre2)-(.7,0)$) node[left] {deleted} -- ($(pre2.west)-(4.5mm,0)$);
    	\draw[-latex, rounded corners=1] ($(pre6)-(4.5mm,0)$) --++(-2.5mm,0mm) -- node[pos=1, left] {replaces} ($(pre4)-(7mm,0)$) -- ($(pre4.west)-(4.5mm,0)$);
    	\draw[-latex, rounded corners=1] ($(pre7)-(4.5mm,0)$) --++(-3.5mm,0mm) -- node[midway, left] {replaces} ($(pre5)-(8mm,0)$) -- ($(pre5.west)-(4.5mm,0)$);
    	\draw[-latex, draw=gitGreenFull] ($(pre8)-(.7,0)$) node[left] {added} -- ($(pre8.west)-(4.5mm,0)$);
    	\draw[-latex, rounded corners=1] ($(pre12)-(4.5mm,0)$) --++(-2.5mm,0mm) -- node[pos=1, left] {replaces} ($(pre10)-(7mm,0)$) -- ($(pre10.west)-(4.5mm,0)$);
    	\draw[-latex, draw=gitRedFull] ($(pre11)-(.8,0)$) node[left] {deleted} -- ($(pre11.west)-(4.5mm,0)$);
    	\draw[-latex, draw=gitGreenFull] ($(pre14)-(.7,0)$) node[left] {added} -- ($(pre14.west)-(4.5mm,0)$);
    	
    	(post1.east)+(1.75,0)
    	
    	\draw[-latex, draw=gitRedFull] ($(post2.east)+(2.1,0)$) node[right] {deleted} -- ($(post2.east)+(1.85,0)$);
    	\draw[rounded corners=1] ($(post8.east)+(1.85,-1.5mm)$) -| ($(post6.east)+(1.9,1.5mm)$) --($(post6.east)+(1.85,1.5mm)$);
    	\draw[rounded corners=1] ($(post5.east)+(1.85,-1.5mm)$) -| ($(post4.east)+(1.9,1.5mm)$) --($(post4.east)+(1.85,1.5mm)$);
    	\draw[-latex, rounded corners=1] ($(post7.east)+(1.9,0)$) --++(2mm,0mm) -- node[midway, right] {replaces} ($(post4.east)+(2.1,-1.5mm)$) -- ($(post4.east)+(1.9,-1.5mm)$);
    	\draw[rounded corners=1] ($(post11.east)+(1.85,-1.5mm)$) -| ($(post10.east)+(1.9,1.5mm)$) --($(post10.east)+(1.85,1.5mm)$);
    	\draw[-latex, rounded corners=1] ($(post12.east)+(1.85,0)$) --++(2.5mm,0mm) -- node[midway, right] {replaces} ($(post10.east)+(2.1,-1.5mm)$) -- ($(post10.east)+(1.9,-1.5mm)$);
    	
    	\draw[-latex, draw=gitGreenFull] ($(post14.east)+(2.1,0)$) node[right] {added} -- ($(post14.east)+(1.85,0)$);
        	
    	\node[anchor=east] at ($(pre1.west)+(-.7,.2)$) {\textbf{line based}};
    	\node[anchor=west] at ($(post1.east)+(2.1,.2)$) {\textbf{block based}};
    \end{tikzpicture}
    \caption{Identification of replacements using line- and block-based analysis.}
    \label{fig:git_line_block}
\end{figure}
In the discussion above, we have considered a purely line-based approach, which treats every modified line of code as a separate entity.
However, it is common that developers edit contiguous regions of code, consisting of multiple adjacent lines, with a single modification.
As illustrated in Figure \ref{fig:git_line_block}, \texttt{git2net} therefore provides an option to analyse co-edits at the granularity of such contiguous code regions rather than lines. 
Compared to previous approaches, which have used programming language constructs like functions to identify co-edits at a granularity smaller than files~\cite{Joblin2015}, this approach has the advantage that it is agnostic of the programming language.
It further allows to analyse co-edit relations in files that do not represent source code, e.g. in text documents.

To explain our approach of identifying edited \emph{blocks} of code, we distinguished between different cases contained in Figure \ref{fig:git_line_block}:
For deleted lines (e.g. line 2 in Fig. \ref{fig:git_line_block}) a normal co-editing relationship is recorded. As the effort required for deletions can vary both between projects and the type of analysis performed, we mark these cases in the database but do not specify a Levenshtein edit distance.
Edits exclusively consisting of added lines are recorded in the database but not considered as co-edits (neither by a line-based nor by a block-based approach) as no previous author exists. The Levenshtein edit distance for pure additions matches the number of characters that were added.
For cases where a set $D$ of deleted lines is replaced by a set $A$ of added lines, the line-based approach matches each line $d_i \in D$ with a line $a_i \in A$ for $i \leq \min(\vert D \vert, \vert A \vert)$.
If $\vert D \vert < \vert A \vert$, a line-based approach would thus treat the excess lines in $A$ as added lines, thus not considering them as a co-edit.
This is the case in line 4-5 in Fig.~\ref{fig:git_line_block}. 
With our block-based approach, we instead identify that a block of lines (lines 4-5) in the original file is replaced by a new block (lines 3-5) in the new file.
If $\vert D \vert > \vert A \vert$, a line-based approach identifies the excess lines in $D$ as deleted lines (see line 7-8 in Fig.~\ref{fig:git_line_block}).
Through a block-based analysis we are instead able to identify that a block of lines (lines 7-8) in the original file is replaced by a new block of lines (line 7) in the new file.

While for the line-based approach, all editing statistics such as the Levenshtein edit distance or the entropy are computed on pairs of lines $(d_i, a_i)$, the block based approach considers the set of lines in $A$ as a replacement of the lines contained in $D$.
Consequently all statistics are computed for the pair of code blocks $(D,A)$.

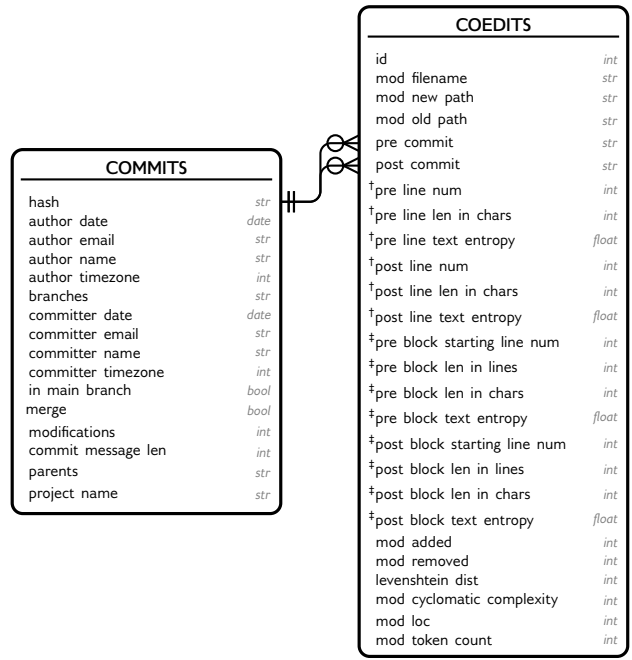
\begin{figure}[t!]
\begin{center}
\begin{tikzpicture}[every node/.style={font=\sffamily}, node distance=5cm]
   	\matrix  [entity=commits]  {
   		\entitynamenode
   		\str{hash}{ }
   		\date{author date}{ }
   		\str{author email}{ }
   		\str{author name}{ }
   		\int{author timezone}{ }
   		\str{branches}{ }
   		\date{committer date}{ }
   		\str{committer email}{ }
   		\str{committer name}{ }
   		\int{committer timezone}{ }
   		\bool{in main branch}{ }
   		\bool{merge}{\phantom{i}}
   		\int{modifications}{ }
   		\int{commit message len}{ }
   		\str{parents}{ }
   		\str{project name}{ }
   	};
   	
   	\matrix  [entity=coedits, node distance=1cm, right=of commits]  {
   		\entitynamenode
   		\int{id}{ }
   		\str{mod filename}{ }
   		\str{mod new path}{ }
   		\str{mod old path}{ }
   		\str{pre commit}{ }
   		\str{post commit}{ }
   		\int{pre line num}{\textsuperscript{\dag}}
		\int{pre line len in chars}{\textsuperscript{\dag}}
		\float{pre line text entropy}{\textsuperscript{\dag}}
		\int{post line num}{\textsuperscript{\dag}}
		\int{post line len in chars}{\textsuperscript{\dag}}
		\float{post line text entropy}{\textsuperscript{\dag}}
   		\int{pre block starting line num}{\textsuperscript{\ddag}}
   		\int{pre block len in lines}{\textsuperscript{\ddag}}
   		\int{pre block len in chars}{\textsuperscript{\ddag}}
   		\float{pre block text entropy}{\textsuperscript{\ddag}}
   		\int{post block starting line num}{\textsuperscript{\ddag}}
   		\int{post block len in lines}{\textsuperscript{\ddag}}
   		\int{post block len in chars}{\textsuperscript{\ddag}}
   		\float{post block text entropy}{\textsuperscript{\ddag}}
   		\int{mod added}{ }
   		\int{mod removed}{ }
   		\int{levenshtein dist}{ }
   		\int{mod cyclomatic complexity}{ }
   		\int{mod loc}{ }
   		\int{mod token count}{ }
   	};

	\draw [omany to one] (coedits-pre commit) --++(-2.3,0) |- (commits-hash);
	\draw [omany to one] (coedits-post commit) --++(-2.3,0) |- (commits-hash);
\end{tikzpicture}
\end{center}
\caption{Relations in the co-editing database. Elements marked \dag~only occur for line based analysis, whereas entries marked \ddag~are specific to block based analysis.}
\label{fig:entity_relationship_coediting}
\end{figure}
After evaluating each commit, results are written to an \texttt{sqlite} database.
This allows to pause and resume an analysis at any point in time and helps to prevent data loss from system crashes. 
The resulting database scheme is shown in Figure \ref{fig:entity_relationship_coediting}.

\subsection{From Co-Edits to Networks}
\input{figures/coediting_network.tex}

Given the database of co-editing relationships generated by the approach described above, \texttt{git2net} provides procedures to generate three different types of network projections: (i) co-editing networks, (ii) directed acyclic graphs of edit sequences for a given file, and (iii) bipartite networks linking developers to edited files.

The process of generating co-editing networks is illustrated in an example shown in Figure \ref{fig:coediting_network}. 
The left column shows three developers ($A$, $B$, and $C$) editing three colour-coded files. 
Modified lines are shown in red. 
Edges between files represent the number of overlapping lines, which for illustrative purposes we show instead of the more granular Levenshtein edit distance. 
Given these edges, we generate a temporal network connecting the developers (cf. Fig. \ref{fig:coediting_network}, centre for a time-unfolded representation).
A link $(A, B; t, w)$ in this network represents a commit by developer $A$ at time $t$ in which $w$ lines originally authored by developer $B$ are modified.
By the aggregation of time-stamped links over a (moving) time window we obtain co-editing networks as shown in the right column of Figure \ref{fig:coediting_network}.

Apart from co-editing networks, \texttt{git2net} supports the construction of file-based directed acyclic graphs (DAGs) of commits based on co-editing relationships.
 \begin{figure}[b!]
    \centering
    \begin{tikzpicture}\sffamily
    	\tikzset{every loop/.style={}}
    
    	\coordinate (offset) at (0,.6);
    	\coordinate (distance) at (0,.8);
    	
    	
    	\coordinate (A) at (0,0);
    	\coordinate (B) at (1,0);
    	\coordinate (C) at (2,0);
    
    	\draw[-latex, black!70] ($(B)+(-.25,4.3)$) -- ($(B)-(.25,0)$) node[left] {\footnotesize time};
    	
    	\coordinate (x) at ($(A)+(offset)+4*(distance)$);
    	\draw[black!30, thick, rounded corners, fill=white] ($(x)+(-1.2,-.35)$) -| ($(x)+(2.7,.35)$) -- ($(x)+(-1.2,.35)$) -- cycle;
    	\node[white, circle, draw=none, minimum size=5.5mm, inner sep=0mm, fill=black!70] at ($(x)-(.85,0)$) {\textbf{1}};
    	
    	\coordinate (x) at ($(A)+(offset)+3*(distance)$);
    	\draw[black!30, thick, rounded corners, fill=white] ($(x)+(-1.2,-.35)$) -| ($(x)+(2.7,.35)$) -- ($(x)+(-1.2,.35)$) -- cycle;
    	\node[white, circle, draw=none, minimum size=5.5mm, inner sep=0mm, fill=black!70] at ($(x)-(.85,0)$) {\textbf{2}};
    	
    	\coordinate (x) at ($(A)+(offset)+2*(distance)$);
    	\draw[black!30, thick, rounded corners, fill=white] ($(x)+(-1.2,-.35)$) -| ($(x)+(2.7,.35)$) -- ($(x)+(-1.2,.35)$) -- cycle;
    	\node[white, circle, draw=none, minimum size=5.5mm, inner sep=0mm, fill=black!70] at ($(x)-(.85,0)$) {\textbf{3}};
    	
    	\coordinate (x) at ($(A)+(offset)+(distance)$);
    	\draw[black!30, thick, rounded corners, fill=white] ($(x)+(-1.2,-.35)$) -| ($(x)+(2.7,.35)$) -- ($(x)+(-1.2,.35)$) -- cycle;
    	\node[white, circle, draw=none, minimum size=5.5mm, inner sep=0mm, fill=black!70] at ($(x)-(.85,0)$) {\textbf{4}};
    	
    	\coordinate (x) at ($(A)+(offset)$);
    	\draw[black!30, thick, rounded corners, fill=white] ($(x)+(-1.2,-.35)$) -| ($(x)+(2.7,.35)$) -- ($(x)+(-1.2,.35)$) -- cycle;
    	\node[white, circle, draw=none, minimum size=5.5mm, inner sep=0mm, fill=black!70] at ($(x)-(.85,0)$) {\textbf{5}};
    	
    	
    	\coordinate (x) at ($(A)+(offset)+4*(distance)$);
    	\draw[customB, thick, rounded corners=2, fill=white] ($(x)+(.15,.25)$) -| ($(x)-(.3,.25)$) -| ($(x)+(.3,.1)$);
    	\draw[colorSG, ultra thick,line cap=round] ($(x)+(-.2,.125)$) -- ($(x)+(.2,.125)$);
    	\draw[colorSG, ultra thick,line cap=round] ($(x)+(-.2,0)$) -- ($(x)+(.2,0)$);
    	\draw[colorSG, ultra thick,line cap=round] ($(x)+(-.2,-.125)$) -- ($(x)+(.2,-.125)$);	
    	\draw[customB, thick, rounded corners=2, fill=white] ($(x)+(.15,.25)$) |- ($(x)+(.3,.1)$);
    	\draw[customB, thick, line cap=round] ($(x)+(.15,.25)$) -- ($(x)+(.3,.1)$);
    	
    	\coordinate (x) at ($(A)+(offset)+3*(distance)$);
    	\draw[customB, thick, rounded corners=2, fill=white] ($(x)+(.15,.25)$) -| ($(x)-(.3,.25)$) -| ($(x)+(.3,.1)$);
    	\draw[customB, ultra thick,line cap=round] ($(x)+(-.2,.125)$) -- ($(x)+(.2,.125)$);
    	\draw[colorSG, ultra thick,line cap=round] ($(x)+(-.2,0)$) -- ($(x)+(.2,0)$);
    	\draw[colorSG, ultra thick,line cap=round] ($(x)+(-.2,-.125)$) -- ($(x)+(.2,-.125)$);	
    	\draw[customB, thick, rounded corners=2, fill=white] ($(x)+(.15,.25)$) |- ($(x)+(.3,.1)$);
    	\draw[customB, thick, line cap=round] ($(x)+(.15,.25)$) -- ($(x)+(.3,.1)$);
    	
    	\coordinate (x) at ($(A)+(offset)+2*(distance)$);
    	\draw[customB, thick, rounded corners=2, fill=white] ($(x)+(.15,.25)$) -| ($(x)-(.3,.25)$) -| ($(x)+(.3,.1)$);
    	\draw[colorSG, ultra thick,line cap=round] ($(x)+(-.2,.125)$) -- ($(x)+(.2,.125)$);
    	\draw[colorSG, ultra thick,line cap=round] ($(x)+(-.2,0)$) -- ($(x)+(.2,0)$);
    	\draw[customB, ultra thick,line cap=round] ($(x)+(-.2,-.125)$) -- ($(x)+(.2,-.125)$);	
    	\draw[customB, thick, rounded corners=2, fill=white] ($(x)+(.15,.25)$) |- ($(x)+(.3,.1)$);
    	\draw[customB, thick, line cap=round] ($(x)+(.15,.25)$) -- ($(x)+(.3,.1)$);
    	
    	\coordinate (x) at ($(A)+(offset)$);
    	\draw[customB, thick, rounded corners=2, fill=white] ($(x)+(.15,.25)$) -| ($(x)-(.3,.25)$) -| ($(x)+(.3,.1)$);
    	\draw[customB, ultra thick,line cap=round] ($(x)+(-.2,.125)$) -- ($(x)+(.2,.125)$);
    	\draw[customB, ultra thick,line cap=round] ($(x)+(-.2,0)$) -- ($(x)+(.2,0)$);
    	\draw[colorSG, ultra thick,line cap=round] ($(x)+(-.2,-.125)$) -- ($(x)+(.2,-.125)$);	
    	\draw[customB, thick, rounded corners=2, fill=white] ($(x)+(.15,.25)$) |- ($(x)+(.3,.1)$);
    	\draw[customB, thick, line cap=round] ($(x)+(.15,.25)$) -- ($(x)+(.3,.1)$);
    	
    	
    	\coordinate (x) at ($(B)+(offset)+3*(distance)$);
    	\draw[customY, thick, rounded corners=2, fill=white] ($(x)+(.15,.25)$) -| ($(x)-(.3,.25)$) -| ($(x)+(.3,.1)$);
    	\draw[colorSG, ultra thick,line cap=round] ($(x)+(-.2,.125)$) -- ($(x)+(.2,.125)$);
    	\draw[colorSG, ultra thick,line cap=round] ($(x)+(-.2,0)$) -- ($(x)+(.2,0)$);
    	\draw[colorSG, ultra thick,line cap=round] ($(x)+(-.2,-.125)$) -- ($(x)+(.2,-.125)$);	
    	\draw[customY, thick, rounded corners=2, fill=white] ($(x)+(.15,.25)$) |- ($(x)+(.3,.1)$);
    	\draw[customY, thick, line cap=round] ($(x)+(.15,.25)$) -- ($(x)+(.3,.1)$);
    	
    	\coordinate (x) at ($(B)+(offset)+1*(distance)$);
    	\draw[customY, thick, rounded corners=2, fill=white] ($(x)+(.15,.25)$) -| ($(x)-(.3,.25)$) -| ($(x)+(.3,.1)$);
    	\draw[customY, ultra thick,line cap=round] ($(x)+(-.2,.125)$) -- ($(x)+(.2,.125)$);
    	\draw[customY, ultra thick,line cap=round] ($(x)+(-.2,0)$) -- ($(x)+(.2,0)$);
    	\draw[colorSG, ultra thick,line cap=round] ($(x)+(-.2,-.125)$) -- ($(x)+(.2,-.125)$);	
    	\draw[customY, thick, rounded corners=2, fill=white] ($(x)+(.15,.25)$) |- ($(x)+(.3,.1)$);
    	\draw[customY, thick, line cap=round] ($(x)+(.15,.25)$) -- ($(x)+(.3,.1)$);
    	
    	\coordinate (x) at ($(B)+(offset)$);
    	\draw[customY, thick, rounded corners=2, fill=white] ($(x)+(.15,.25)$) -| ($(x)-(.3,.25)$) -| ($(x)+(.3,.1)$);
    	\draw[colorSG, ultra thick,line cap=round] ($(x)+(-.2,.125)$) -- ($(x)+(.2,.125)$);
    	\draw[customY, ultra thick,line cap=round] ($(x)+(-.2,0)$) -- ($(x)+(.2,0)$);
    	\draw[customY, ultra thick,line cap=round] ($(x)+(-.2,-.125)$) -- ($(x)+(.2,-.125)$);	
    	\draw[customY, thick, rounded corners=2, fill=white] ($(x)+(.15,.25)$) |- ($(x)+(.3,.1)$);
    	\draw[customY, thick, line cap=round] ($(x)+(.15,.25)$) -- ($(x)+(.3,.1)$);
    	
    	
    	\coordinate (x) at ($(C)+(offset)+4*(distance)$);
    	\draw[customG, thick, rounded corners=2, fill=white] ($(x)+(.15,.25)$) -| ($(x)-(.3,.25)$) -| ($(x)+(.3,.1)$);
    	\draw[colorSG, ultra thick,line cap=round] ($(x)+(-.2,.125)$) -- ($(x)+(.2,.125)$);
    	\draw[colorSG, ultra thick,line cap=round] ($(x)+(-.2,0)$) -- ($(x)+(.2,0)$);
    	\draw[colorSG, ultra thick,line cap=round] ($(x)+(-.2,-.125)$) -- ($(x)+(.2,-.125)$);	
    	\draw[customG, thick, rounded corners=2, fill=white] ($(x)+(.15,.25)$) |- ($(x)+(.3,.1)$);
    	\draw[customG, thick, line cap=round] ($(x)+(.15,.25)$) -- ($(x)+(.3,.1)$);
    		
    	\coordinate (x) at ($(C)+(offset)+2*(distance)$);
    	\draw[customG, thick, rounded corners=2, fill=white] ($(x)+(.15,.25)$) -| ($(x)-(.3,.25)$) -| ($(x)+(.3,.1)$);
    	\draw[colorSG, ultra thick,line cap=round] ($(x)+(-.2,.125)$) -- ($(x)+(.2,.125)$);
    	\draw[customG, ultra thick,line cap=round] ($(x)+(-.2,0)$) -- ($(x)+(.2,0)$);
    	\draw[colorSG, ultra thick,line cap=round] ($(x)+(-.2,-.125)$) -- ($(x)+(.2,-.125)$);	
    	\draw[customG, thick, rounded corners=2, fill=white] ($(x)+(.15,.25)$) |- ($(x)+(.3,.1)$);
    	\draw[customG, thick, line cap=round] ($(x)+(.15,.25)$) -- ($(x)+(.3,.1)$);
    	
    	\coordinate (x) at ($(C)+(offset)+(distance)$);
    	\draw[customG, thick, rounded corners=2, fill=white] ($(x)+(.15,.25)$) -| ($(x)-(.3,.25)$) -| ($(x)+(.3,.1)$);
    	\draw[colorSG, ultra thick,line cap=round] ($(x)+(-.2,.125)$) -- ($(x)+(.2,.125)$);
    	\draw[colorSG, ultra thick,line cap=round] ($(x)+(-.2,0)$) -- ($(x)+(.2,0)$);
    	\draw[customG, ultra thick,line cap=round] ($(x)+(-.2,-.125)$) -- ($(x)+(.2,-.125)$);	
    	\draw[customG, thick, rounded corners=2, fill=white] ($(x)+(.15,.25)$) |- ($(x)+(.3,.1)$);
    	\draw[customG, thick, line cap=round] ($(x)+(.15,.25)$) -- ($(x)+(.3,.1)$);
    	
    	\coordinate (x) at ($(C)+(offset)$);
    	\draw[customG, thick, rounded corners=2, fill=white] ($(x)+(.15,.25)$) -| ($(x)-(.3,.25)$) -| ($(x)+(.3,.1)$);
    	\draw[customG, ultra thick,line cap=round] ($(x)+(-.2,.125)$) -- ($(x)+(.2,.125)$);
    	\draw[customG, ultra thick,line cap=round] ($(x)+(-.2,0)$) -- ($(x)+(.2,0)$);
    	\draw[colorSG, ultra thick,line cap=round] ($(x)+(-.2,-.125)$) -- ($(x)+(.2,-.125)$);	
    	\draw[customG, thick, rounded corners=2, fill=white] ($(x)+(.15,.25)$) |- ($(x)+(.3,.1)$);
    	\draw[customG, thick, line cap=round] ($(x)+(.15,.25)$) -- ($(x)+(.3,.1)$);
    	
    	
    	\footnotesize
    	\path[-latex, customB] ($(A)+(offset)+3*(distance)+(.3,0)$) edge[bend right=15] node[pos=.75, right, xshift=-.3mm] {2} ($(A)+(offset)+4*(distance)+(.3,0)$);
    	\path[-latex, customB] ($(A)+(offset)+2*(distance)-(.3,0)$) edge[bend left=15] node[pos=.5, right] {} ($(A)+(offset)+4*(distance)-(.3,0)$);
    	\path[-latex, customB] ($(A)+(offset)+2*(distance)+(.3,0)$) edge[bend right=15] node[pos=.5, right] {} ($(A)+(offset)+3*(distance)+(.3,0)$);
    	\path[-latex, customB] ($(A)+(offset)+0*(distance)+(.3,0)$) edge[bend right=15] node[pos=.5, right] {} ($(A)+(offset)+3*(distance)+(.3,0)$);
    	
    	\path[-latex, customY] ($(B)+(offset)+1*(distance)-(.3,0)$) edge[bend left=15] node[pos=.5, right] {} ($(B)+(offset)+3*(distance)-(.3,0)$);
    	\path[-latex, customY] ($(B)+(offset)+0*(distance)+(.3,0)$) edge[bend right=15] node[pos=.5, right] {} ($(B)+(offset)+3*(distance)+(.3,0)$);
    	
    	\path[-latex, customG] ($(C)+(offset)+2*(distance)+(.3,0)$) edge[bend right=15] node[pos=.5, right, xshift=-.6mm] {2} ($(C)+(offset)+4*(distance)+(.3,0)$);
    	\path[-latex, customG] ($(C)+(offset)+1*(distance)+(.3,0)$) edge[bend right=15] node[pos=.5, right] {} ($(C)+(offset)+2*(distance)+(.3,0)$);
    	\path[-latex, customG] ($(C)+(offset)+1*(distance)-(.3,0)$) edge[bend left=15] node[pos=.5, right] {} ($(C)+(offset)+4*(distance)-(.3,0)$);
    	\path[-latex, customG] ($(C)+(offset)+0*(distance)+(.3,0)$) edge[bend right=15] node[pos=.5, right] {} ($(C)+(offset)+2*(distance)+(.3,0)$);

    	\normalsize
    	
    	\coordinate (x) at (2.8,0);
    	\coordinate (y) at ($4*(distance)+(0,.7)$);
    	\draw[black!70, rounded corners=3, thick] ($(x)+(offset)-(0,.35)$) -| ($(x)+(offset)+.5*(y)+(.2,-.35)$) -- ($(x)+(offset)+.5*(y)+(.4,-.35)$) -- ($(x)+(offset)+.5*(y)+(.2,-.35)$) |- ($(x)+(offset)+(y)-(0,.35)$);
    		
    	
    	\node[white, circle, draw=none, minimum size=5.5mm, inner sep=0mm, fill=customB] (B1) at ($(A)+(offset)+4*(distance)+(4.6,0)$) {\textbf{1}};
    	\node[white, circle, draw=none, minimum size=5.5mm, inner sep=0mm, fill=customB] (B2) at ($(A)+(offset)+3.5*(distance)+(3.6,0)$) {\textbf{2}};
    	\node[white, circle, draw=none, minimum size=5.5mm, inner sep=0mm, fill=customB] (B3) at ($(A)+(offset)+3.5*(distance)+(5.6,0)$) {\textbf{3}};
    	\node[white, circle, draw=none, minimum size=5.5mm, inner sep=0mm, fill=customB] (B5) at ($(A)+(offset)+3*(distance)+(4.6,0)$) {\textbf{5}};
    	
    	\node[white, circle, draw=none, minimum size=5.5mm, inner sep=0mm, fill=customY] (Y2) at ($(A)+(offset)+2*(distance)+(4.6,0)$) {\textbf{2}};
    	\node[white, circle, draw=none, minimum size=5.5mm, inner sep=0mm, fill=customY] (Y4) at ($(A)+(offset)+2*(distance)+(3.6,0)$) {\textbf{4}};
    	\node[white, circle, draw=none, minimum size=5.5mm, inner sep=0mm, fill=customY] (Y5) at ($(A)+(offset)+2*(distance)+(5.6,0)$) {\textbf{5}};
    	
    	\node[white, circle, draw=none, minimum size=5.5mm, inner sep=0mm, fill=customG] (G1) at ($(A)+(offset)+1*(distance)+(4.6,0)$) {\textbf{1}};
    	\node[white, circle, draw=none, minimum size=5.5mm, inner sep=0mm, fill=customG] (G3) at ($(A)+(offset)+.5*(distance)+(3.6,0)$) {\textbf{3}};
    	\node[white, circle, draw=none, minimum size=5.5mm, inner sep=0mm, fill=customG] (G4) at ($(A)+(offset)+.5*(distance)+(5.6,0)$) {\textbf{4}};
    	\node[white, circle, draw=none, minimum size=5.5mm, inner sep=0mm, fill=customG] (G5) at ($(A)+(offset)+0*(distance)+(4.6,0)$) {\textbf{5}};
    	
    	
    	\footnotesize
    
    	\draw[-latex, customB] (B5) -- (B2);
    	\draw[-latex, customB] (B3) -- (B2);
    	\draw[-latex, customB] (B3) -- (B1);
    	\draw[-latex, customB] (B2) -- node[pos=.35,above] {2} (B1);
    	
    	\draw[-latex, customY] (Y5) -- (Y2);
    	\draw[-latex, customY] (Y4) -- (Y2);
    	
    	\draw[-latex, customG] (G5) -- (G3);
    	\draw[-latex, customG] (G4) -- (G3);
    	\draw[-latex, customG] (G4) -- (G1);
    	\draw[-latex, customG] (G3) -- node[pos=.35,above] {2} (G1);
    			
    \end{tikzpicture}
    \caption{Process of creating file based directed acyclic co-editing graphs. The left hand side shows a set of commits modifying three colour coded files. For each file a directed acyclic graph is generated linking consecutive commits with overlapping changes.}
     \label{fig:dag}
 \end{figure}
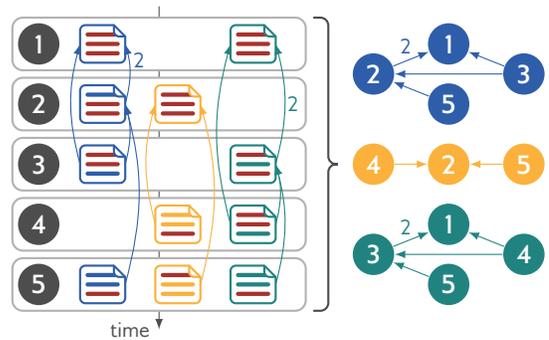
Each path in this DAG represent a sequence of consecutive co-editing relationships of developers editing the given file, i.e. a sequence of commits containing file modifications that built upon each other.
The nodes in this graph represent commits and edges represent co-editing relationships between the authors of the commits.
An example for the construction of such a DAG from a set of five commits containing file modifications is shown in Figure \ref{fig:dag}. 
Individual connected components of the DAG represent proxies of knowledge flow for this file.
This has been highly valuable in our own research as it immediately allows the extraction of paths from the co-editing relationships.
Analysing these paths with the methods provided by the software package \texttt{pathpy}~\cite{pathpy} allows to trace knowledge flow within specific areas of the development---a topic we identified as highly relevant in discussions with practitioners from software development companies.

To additionally facilitate coarse-grained analyses at the level of file-based coauthorship relations, \texttt{git2net} finally supports the construction of bipartite file-developer networks, where directed links $(d,f) \in D \times F$ indicate that a developer $d \in D$ has modified a file $f \in F$.

\subsection{Usage of git2net}

\texttt{git2net} comes as a \texttt{python} package that can be installed via the \texttt{python} package manager \texttt{pip}.
During the installation all dependencies, which consist of the \texttt{python} packages \texttt{pandas}, \texttt{python\_Levenshtein}, \texttt{pyDriller}, \texttt{progressbar2}, and \texttt{pathpy}, will be installed automatically.
\texttt{git2net} runs on all major operating systems and has been tested under Windows, Mac OS X, and Linux.
Assuming that the \texttt{git} repository that shall be examined has been cloned to a directory \texttt{repo}, our tool can be launched by the command
\begin{shaded}
\texttt{./git2net.py mine repo coedits.db}
\end{shaded}
where \texttt{coedits.db} indicates the \texttt{sqlite} database file where the results will be stored.
An optional parameter \texttt{--exclude} can be used to pass a text file that contains paths of files or directories in the repository tree that shall be excluded from the analysis.
In our own analyses of a large commercial software project, this function has proven crucial to exclude directories containing large binary files or external Open Source software dependencies that would considerably distort the analysis.
While the analysis of co-edited code uses the line-based approach described above by default, an optional command line switch \texttt{--use-blocks} can be used to use the block-based extraction of co-editing relations instead.

In addition to the command line interface outlined above, \texttt{git2net} provides an API that can be used for the development of custom repository mining scripts. 
In particular, the API provides methods that allow to extract co-edit relations from individual commits that can be passed as \texttt{PyDriller} objects. 
It can further be used to augment the analysis of edited code blocks by advanced text mining and code analysis techniques.
In order to generate network projections based on a database of co-edits, \texttt{git2net} can be launched with the command
\begin{shaded}
\texttt{./git2net.py graph [type] coedits.db graph.csv}
\end{shaded}
where \texttt{type} can be \texttt{--coedit}, \texttt{--bipartite}, or \texttt{--dag}.
Depending on the choice, \texttt{git2net} generates a projection of the co-editing database in terms of a temporal co-editing network (cf. Fig.~\ref{fig:coediting_network}), a bipartite network linking authors to files, or a directed acyclic co-editing graph (cf. Fig.~\ref{fig:dag}) respectively.

All networks can be exported in a csv-based format that can be read by popular network analysis packages like \texttt{igraph}~\cite{csardi2006}, \texttt{graphtool}\footnote{\url{https://graph-tool.skewed.de/}}, \texttt{Gephi}~\cite{bastian2009gephi}, and NetworkX~\cite{hagberg2008exploring}.
Time-stamped co-editing networks can further be exported in a format that can be read by the dynamic network analysis and visualisation packages ORA~\cite{carley2012dynamic} and \texttt{pathpy}~\cite{pathpy} via the provided API.
Moreover, all networks can be exported in terms of dynamic and interactive \texttt{d3js} visualisations, which directly run in any \texttt{HTML5}-compliant browser.

\subsection{Experimental Evaluation of Scalability}
\begin{figure}[b!]
    \centering
    \begin{tikzpicture}\sffamily\footnotesize
    \begin{axis}[
    	height=5cm,
    	width=9cm,
    	xmode=log,
    	ymode=log,
    	xmin=0.8, xmax=40,
    	xtick={1,2,4,8,16,32},
    	xticklabels={1,2,4,8,16,32},
    	ytick={59.427777777777784,30,15,5,3.1,1.8571180555555558},
    	yticklabels={59:26,30,15,5,3:06, 1:51},
    	ymin=1.8571180555555558, ymax=65,
    	tick align=outside,
    	tick pos=left,
    	axis x line*=bottom,
    	axis y line*=left,
    	xlabel=number of threads,
    	ylabel=time in minutes,
    ]
    \addplot[thin, black!30, mark=none, dotted]
	    table [row sep=\\]{%
	    	0.1 59.427777777777784 \\
	    	50 59.427777777777784 \\
	    	};
   	\addplot[draw=black!30, mark=none, dotted]
   	table [row sep=\\]{%
   		0.1 5 \\
   		50 5 \\
   	};
	\addplot[draw=black!30, mark=none, dotted]
	table [row sep=\\]{%
		0.1 3.1 \\
		50 3.1 \\
	};
    \addplot[draw=black!30, mark=none]
    	table {
    		1 59.427777777777784
    		2 29.713888888888892
    		3 19.80925925925926
    		4 14.856944444444446
    		5 11.885555555555557
    		6 9.90462962962963
    		7 8.48968253968254
    		8 7.428472222222223
    		9 6.603086419753087
    		10 5.942777777777779
    		11 5.402525252525253
    		12 4.952314814814815
    		13 4.571367521367522
    		14 4.24484126984127
    		15 3.9618518518518524
    		16 3.7142361111111115
    		17 3.4957516339869286
    		18 3.3015432098765434
    		19 3.127777777777778
    		20 2.9713888888888893
    		21 2.8298941798941804
    		22 2.7012626262626265
    		23 2.5838164251207734
    		24 2.4761574074074075
    		25 2.377111111111111
    		26 2.285683760683761
    		27 2.2010288065843624
    		28 2.122420634920635
    		29 2.0492337164750962
    		30 1.9809259259259262
    		31 1.917025089605735
    		32 1.8571180555555558
    	};
    
    \addplot [color=customG, only marks,mark=0,]
    	plot [error bars/.cd, y dir = both, y explicit]
    	table[row sep=crcr, x index=0, y index=1, y error index=2]{
    		1 59.427777777777784 0.1454558606842888 \\
    		2 29.305555555555554 0.05443310539518299 \\
    		3 19.66111111111111 0.03788383804718221 \\
    		4 14.799999999999999 0.03118047822311583 \\
    		5 11.877777777777778 0.02965855070008654 \\
    		6 9.938888888888888 0.00680413817439733 \\
    		7 8.572222222222221 0.018002057495577053 \\
    		8 7.527777777777778 0.013608276348795386 \\
    		9 6.716666666666666 7.691850745534255e-16 \\
    		10 6.088888888888889 0.006804138174397693 \\
    		11 5.577777777777778 0.013608276348795384 \\
    		12 5.172222222222223 0.006804138174397692 \\
    		13 4.833333333333333 0.01178511301977575 \\
    		14 4.561111111111111 0.006804138174397694 \\
    		15 4.333333333333333 0.0 \\
    		16 4.133333333333334 0.0 \\
    		17 3.9499999999999997 0.02041241452319326 \\
    		18 3.7999999999999994 0.01178511301977575 \\
    		19 3.6666666666666665 0.02041241452319326 \\
    		20 3.5777777777777775 0.018002057495577498 \\
    		21 3.505555555555555 0.013608276348795384 \\
    		22 3.4333333333333336 0.011785113019775908 \\
    		23 3.3666666666666667 0.01178511301977575 \\
    		24 3.2944444444444443 0.006804138174397693 \\
    		25 3.238888888888889 0.013608276348795386 \\
    		26 3.2055555555555557 0.006804138174397693 \\
    		27 3.1777777777777776 0.006804138174397693 \\
    		28 3.1555555555555554 0.02453266907313282 \\
    		29 3.1277777777777778 0.018002057495577328 \\
    		30 3.1277777777777778 0.018002057495577328 \\
    		31 3.1055555555555556 0.006804138174397693 \\
    		32 3.1 0.0 \\
    	};
    \end{axis}
    \end{tikzpicture}
    \caption{Time required to analyse the \texttt{git} repository of the software package \texttt{igraph}~\cite{csardi2006} for different numbers of parallel processing threads. Both axes are logarithmic. Bars show the mean and standard deviation of three runs. The grey line shows a perfect linear scaling based on the time required by a single-threaded analysis.}
    \label{fig:scalability}
\end{figure}
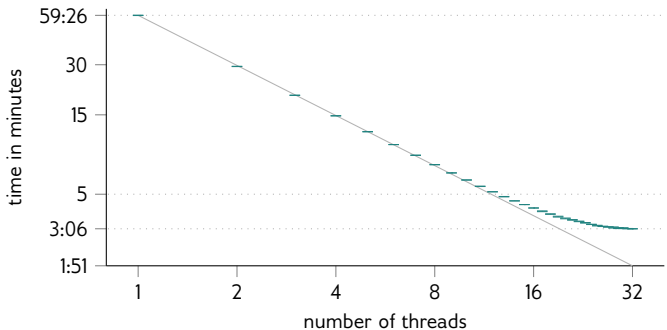

We conclude this section by an experimental evaluation of the scalability of \texttt{git2net}.
In particular, our tool facilitates the analysis of large repositories thanks to the automatic utilisation of multiple processing cores.
By default, \texttt{git2net} uses all available processing core, creating multiple child processes that extract co-edits from independent commits in parallel. 
Through an optional command line switch \texttt{--no-parallel}, multi-core processing can be deactivated.
An optional command line parameter \texttt{--numprocesses N} further allows to limit multi-core processing to at most $N$ processing cores.
Similarly, the API exposed by \texttt{git2net} provides parameters that can be used to control multi-core processing.

In order to evaluate the scalability gains provided by the parallel processing model, we performed an experiment using real-world data.
We specifically cloned the \texttt{git} repository of the Open Source software \texttt{igraph}~\cite{igraph} and used \texttt{git2net} to extract line-based co-editing relationships. 
We then measured the time needed to analyse the full \texttt{git} history with close to 6,000 commits and approximately 35,000 file edits over a period of 14 years.
We repeated this experiment multiple times, using different numbers of processing cores on a recent 16 core desktop processor\footnote{Intel\textregistered~Core\texttrademark~i9 7960X, 16C/32T, 2.80GHz base, 4.2GHz boost}.

Figure \ref{fig:scalability} shows the time required to extract all co-editing relationships from the repository of \texttt{igraph} (y-axis) plotted against the number of processing threads (x-axis).
Up to the number of physical processing cores of the machine (16) we observe an almost perfect linear scaling of processing time, cutting down processing time from close to one hour (single-threaded) to less than 5 minutes. 
Starting from 16 processing cores we observe deviations from the linear scaling that are likely due to the synchronised writing to the \texttt{sqlite} database.
This deviation from the linear scaling is naturally intensified as we exceed the number of physical processing cores, additionally utilising logical cores exposed through Intel's implementation of HW-based multi-threading.

\section{Exemplary Co-Editing Analysis of an Open Source and Commercial Project}
\label{sec:results}

Having discussed the implementation, usage, and scalability of our tool, we now demonstrate its usefulness through four short exemplary studies of real-world software projects.
We apply \texttt{git2net} to (i) the \texttt{gitHub} repository of the Open Source network analysis software \texttt{igraph}~\cite{csardi2006}, and (ii) a large \texttt{git} repository of a commercial software project obtained via an industry collaboration with the software company \textsc{Genua}.
We specifically demonstrate (A) the construction of different static network projections capturing co-editing, co-authorship, and code-ownership relations, (B) a comparative study of fine-grained co-editing networks vs. coarse-grained co-authorship networks generated at the level of files, (C) the analysis of dynamic co-editing networks by means of temporal network analysis techniques, and (D) a comparison of temporal co-editing patterns between an Open Source and a commercial software project. 
These case studies should be seen as seeds for future work that demonstrate the usefulness of our approach rather than as conclusive analyses.
To support such future studies, the co-editing relationships extracted from the Open Source project \texttt{igraph} are available on \texttt{zenodo.org}~\cite{Gote2019}.

\subsection{Static Network Projections}
\label{sec:results:networks}
\begin{figure*}[!htb]
    \centering
    \begin{tikzpicture}\sffamily\footnotesize
        \coordinate (a) at (0,0);
        \coordinate (b) at (4,0);
        \coordinate (c) at (12,0);
        \node[anchor=north west] at ($(a)+(0,-.06\textwidth)$) {\includegraphics[height=.20\textwidth]{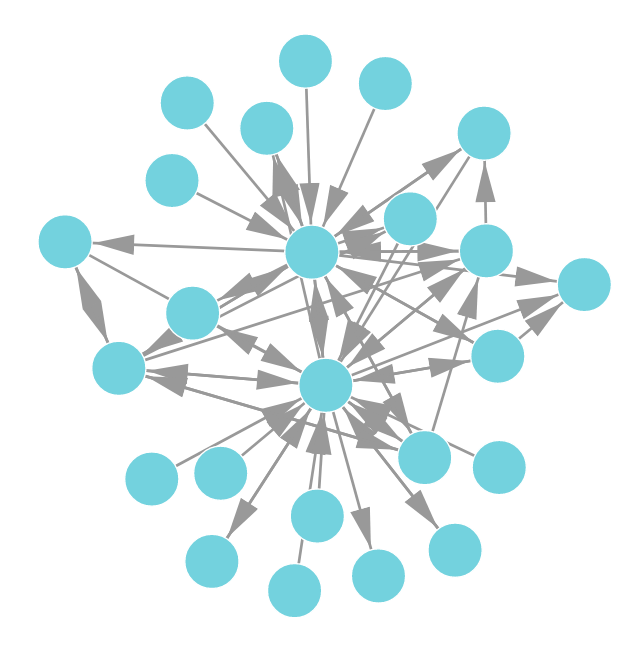}};
        \node[anchor=north west] at (b) {\includegraphics[height=.32\textwidth]{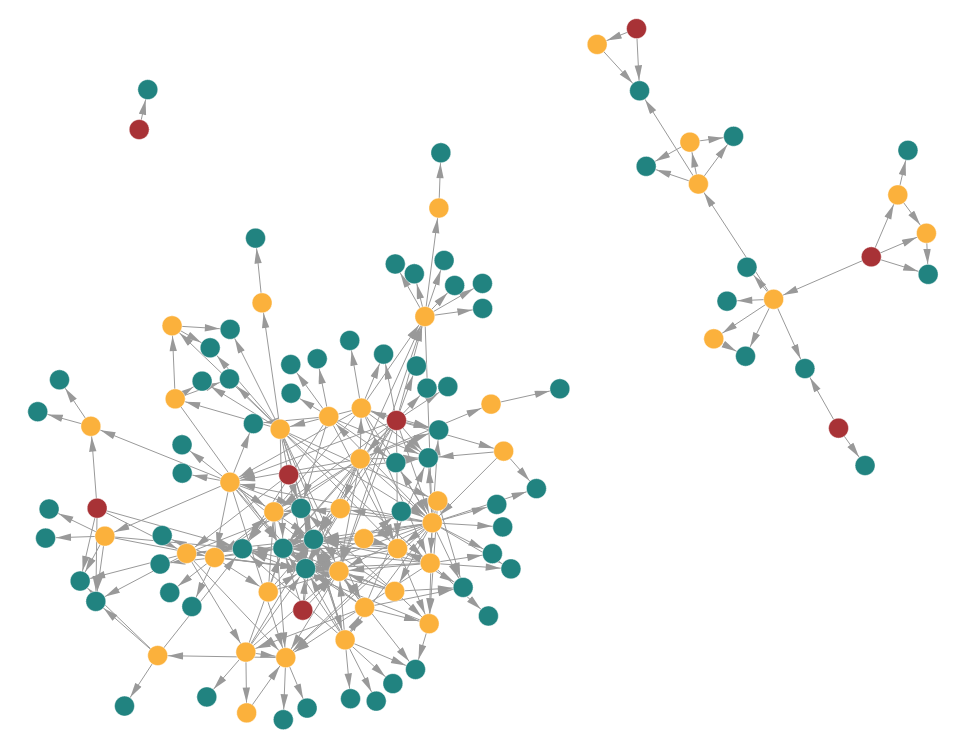}};
        \node[anchor=north west] at (c) {\includegraphics[height=.3\textwidth]{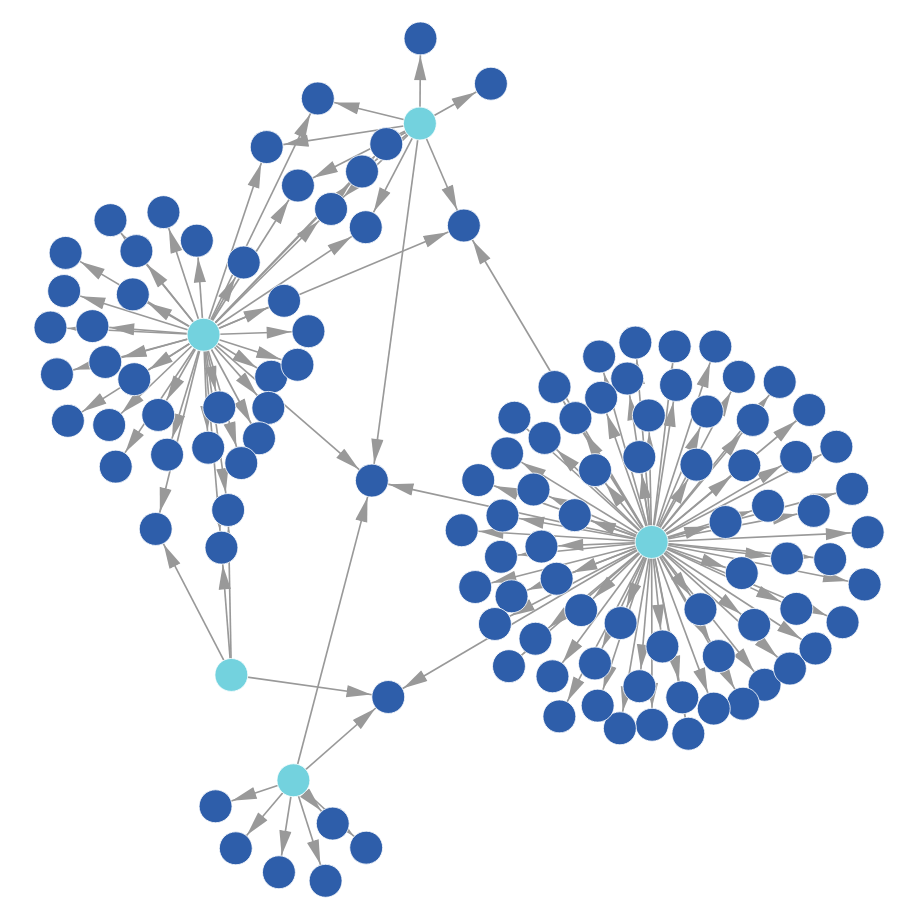}};
        \node at (a) {\textbf{a}};
        \node at (b) {\textbf{b}};
        \node at (c) {\textbf{c}};
    \end{tikzpicture}
    \caption{Three examples for time-aggregated collaboration networks generated by \texttt{git2net} based on co-editing relations in \texttt{igraph} project: \textbf{\sffamily a} shows a time-aggregated, static, directed network of co-editing relations. Each node represents one developer, while a directed link $(A,B)$ indicates that at some point in the development history developer $A$ edited at least one line of code previously written by developer $B$. \textbf{\sffamily b} shows a directed acyclic graph of edits of the source code file \texttt{flow.c}. Nodes represent commits by developers. Root nodes with in-degree zero are marked in red, leaf nodes with out-degree zero are marked in green, intermediary nodes are marked in red. \textbf{\sffamily c} shows a bipartite network linking developers (lightblue) to the files that they edited (blue).
    }
    \label{fig:networks}
\end{figure*}

To demonstrate our tool, we illustrate the three different network projections introduced in \ref{sec:method}, using the co-edit information extracted from the public \texttt{git} repository of the network analysis package \texttt{igraph}~\cite{igraph}.
The resulting networks are shown in Figure \ref{fig:networks}.

Figure~\ref{fig:networks}a shows a static co-editing network where nodes represent developers.
For this initial demonstration we employ a time-aggregated projection, i.e. we use time-stamped co-editing relations $(v,w;t)$ capturing that at time $t$ a developer $v$ edited code originally written by developer $w$ to construct a time-aggregated graph $G(V,E)$ where $(v,w) \in E$ iff $\exists \tau: (v,w;\tau)$.
The directionality of links in this projection allows us to distinguish between team members with different roles:
Nodes with zero in-degree, i.e. developers with no incoming co-edit relations, have never contributed code that was subsequently revised by other developers.
Nodes with zero out-degree, i.e. developers with no outgoing co-edit relations, have never revised code that was originally authored by other developers.
Such a maximally simple static projection can thus give a first ``birds-eye'' view of the collaboration and coordination structures in a software developing team.
It highlights pairs of developers who exhibit strong mutual co-editing relations as well as pairs of developers working independently.
This analysis can be refined by taking into account the time stamps of co-editing events, which we will do in section \ref{sec:results:temporal}.
In section \ref{sec:results:coauthorship} we further discuss the difference between file-based coauthorship networks considered in prior works and the static projection of a fine-grained line-based definition.

Apart from co-editing relations between developers, in section \ref{sec:method} we have argued that \texttt{git2net} also provides a new perspective on the history of commits modifying a \emph{given} file in the repository.
In particular, this information can be used to a construct a directed acyclic graph of commits, where a link $(v,w)$ in the graph indicates that commit $w$ edited a region of source code originally contributed in commit $v$.
Hence, each path from a root node $r$ to a leaf node $l$ in the resulting directed acyclic graph can be interpreted as a time-ordered sequence of commits that transforms code originally introduced in commit $r$ into the ``final'' version contained in $l$.
We highlight that this projection is different from commonly studied commit graphs, which link each commit to their parent commit independent of whether there is an overlap in the edited code.
Fig.~\ref{fig:networks}b illustrates this idea.
It shows the directed acyclic graph of commits for the source code file \texttt{flow.c} in \texttt{igraph}~\cite{igraph}.
Root nodes (with in-degree zero) in which the original version of a region of source code was committed are shown in red, while the commits containing the ``final'' version of code regions (out-degree zero) are highlighted in green.
Intermediary nodes (yellow) represent commits that have both (a) edited code originally contributed in a previous commit and (b) contributed new code that is being revised in a subsequent commit.
The analysis of such directed acyclic graphs can give insights into the complexity of code edits and their distribution across the team or across time.
They further provide a novel abstraction that can be useful for the comparison of software artefacts, development processes, or projects.

In order to make it easy to reproduce file-based definitions of co-authorships used in the literature, \texttt{git2net} finally supports the construction of networks linking developers with the files that they have edited.
The time-aggregated bipartite network resulting from the file edits made in the year 2016 for the project \texttt{igraph} is shown in Fig.~\ref{fig:networks}c.
Apart from being a basis for the construction of file-based coauthorship networks, this simple representation can give a coarse-grained view of code ownership and the distribution of contributions across the development team.

\subsection{Co-editing vs. co-authorship networks}
\label{sec:results:coauthorship}
\begin{figure*}[t!]
    \centering
    \begin{tikzpicture}\footnotesize\sffamily
    	\matrix[ampersand replacement=\&, anchor=east] (table) at (-1.9,.75) {
    		\node[anchor=east] {
    			\begin{tabular}{llccc}
    				& \textbf{network} & \textbf{nodes} & \textbf{edges} & \textbf{$\delta$}\\\midrule
    				\multirow{2}{*}{Open Source} & co-authorship & 24 & 67 & \multirow{2}{*}{1.22}\\
    				& co-editing & 24 & 55 &\\
    				\cmidrule(lr){1-5}
    				\multirow{2}{*}{commercial} & co-authorship & 84 & 2551 & \multirow{2}{*}{2.11}\\
    				& co-editing & 84 & 1211 &
    			\end{tabular}
    		};\\
    	};
    	\begin{axis}[
at={(0,.062\textwidth)},
width=.54\textwidth,
height=.18\textwidth,
xmin=1308830052, xmax=1544702052,
tick align=outside,
tick pos=left,
ylabel style={align=center, yshift=-3mm},
ylabel={Open Source\\[2mm]$\delta$},
x grid style={white!69.01960784313725!black},
y grid style={white!69.01960784313725!black},
axis y line*=left,
axis x line*=bottom,
xtick = {1136073600, 1199145600, 1262304000, 1325376000, 1388534400, 1451606400, 1514764800},
xticklabels = {2006, 2008, 2010, 2012, 2014, 2016, 2018},
scaled ticks=false,
ytick = {0, 1, 2, 3},
yticklabels = {0, 1, 2, 3},
]
\addlegendimage{no markers, customG}
\addplot [thin, black!30, dotted]
table [row sep=\\]{%
1114430052	1 \\
1544702052	1 \\
};
\addplot [thick, customG]
table [row sep=\\]{%
1114430052	1 \\
1117022052	1 \\
1119614052	1 \\
1122206052	1 \\
1124798052	1 \\
1127390052	1 \\
1129982052	1 \\
1132574052	1 \\
1135166052	1 \\
1137758052	1 \\
1140350052	1 \\
1142942052	1 \\
1145534052	1 \\
1148126052	1 \\
1150718052	1 \\
1153310052	1 \\
1155902052	1 \\
1158494052	1 \\
1161086052	1 \\
1163678052	1 \\
1166270052	1 \\
1168862052	1 \\
1171454052	1 \\
1174046052	1 \\
1176638052	1 \\
1179230052	1 \\
1181822052	1 \\
1184414052	1 \\
1187006052	1 \\
1189598052	1 \\
1192190052	1 \\
1194782052	1 \\
1197374052	1 \\
1199966052	1 \\
1202558052	1 \\
1205150052	1 \\
1207742052	1 \\
1210334052	1 \\
1212926052	1 \\
1215518052	1 \\
1218110052	1 \\
1220702052	1 \\
1223294052	1 \\
1225886052	1 \\
1228478052	1 \\
1231070052	1 \\
1233662052	1 \\
1236254052	1 \\
1238846052	1 \\
1241438052	1 \\
1244030052	1 \\
1246622052	1 \\
1249214052	1 \\
1251806052	1 \\
1254398052	1 \\
1256990052	1 \\
1259582052	1 \\
1262174052	1 \\
1264766052	1 \\
1267358052	1 \\
1269950052	1 \\
1272542052	1 \\
1275134052	1 \\
1277726052	1 \\
1280318052	1 \\
1282910052	1 \\
1285502052	1 \\
1288094052	1 \\
1290686052	1 \\
1293278052	1 \\
1295870052	1 \\
1298462052	1 \\
1301054052	1 \\
1303646052	1 \\
1306238052	1 \\
1308830052	1.66666666666667 \\
1311422052	1.2 \\
1314014052	1.2 \\
1316606052	0.6 \\
1319198052	1 \\
1321790052	0.75 \\
1324382052	0.75 \\
1326974052	0.75 \\
1329566052	0.666666666666667 \\
1332158052	0.75 \\
1334750052	0.75 \\
1337342052	0.75 \\
1339934052	1.33333333333333 \\
1342526052	1 \\
1345118052	1 \\
1347710052	1 \\
1350302052	1.33333333333333 \\
1352894052	1 \\
1355486052	1 \\
1358078052	1 \\
1360670052	1.5 \\
1363262052	1 \\
1365854052	1 \\
1368446052	1 \\
1371038052	1 \\
1373630052	1.33333333333333 \\
1376222052	1.5 \\
1378814052	1 \\
1381406052	1 \\
1383998052	3 \\
1386590052	0.6 \\
1389182052	0.6 \\
1391774052	0.333333333333333 \\
1394366052	0.75 \\
1396958052	1 \\
1399550052	0.833333333333333 \\
1402142052	0.8 \\
1404734052	0.666666666666667 \\
1407326052	1 \\
1409918052	1 \\
1412510052	0 \\
1415102052	0.5 \\
1417694052	0.5 \\
1420286052	0.428571428571429 \\
1422878052	0.428571428571429 \\
1425470052	0.166666666666667 \\
1428062052	1 \\
1430654052	0.5 \\
1433246052	0.5 \\
1435838052	0.333333333333333 \\
1438430052	1 \\
1441022052	1 \\
1443614052	1.5 \\
1446206052	0.75 \\
1448798052	0.75 \\
1451390052	0.5 \\
1453982052	0.333333333333333 \\
1456574052	0.333333333333333 \\
1459166052	0.666666666666667 \\
1461758052	0.666666666666667 \\
1464350052	0.666666666666667 \\
1466942052	0.333333333333333 \\
1469534052	0.166666666666667 \\
1472126052	0.3 \\
1474718052	0.3 \\
1477310052	0.142857142857143 \\
1479902052	0 \\
1482494052	0 \\
1485086052	0 \\
1487678052	1 \\
1490270052	0 \\
1492862052	0 \\
1495454052	0.4 \\
1498046052	0.25 \\
1500638052	0.142857142857143 \\
1503230052	0.6 \\
1505822052	0.166666666666667 \\
1508414052	0.4 \\
1511006052	0 \\
1513598052	0 \\
1516190052	0.2 \\
1518782052	0.2 \\
1521374052	0.333333333333333 \\
1523966052	0.5 \\
1526558052	0.333333333333333 \\
1529150052	0.4 \\
1531742052	0.4 \\
1534334052	0.5 \\
1536926052	0 \\
1539518052	0.2 \\
1542110052	0.166666666666667 \\
1544702052	0.3 \\
};
\end{axis}
    	\begin{axis}[
at={(0,-.062\textwidth)},
width=.54\textwidth,
height=.18\textwidth,
xmin=1089054280, xmax=1542654280,
tick align=outside,
tick pos=left,
ylabel style={align=center, yshift=-3mm},
ylabel={commercial\\[2mm]$\delta$},
xtick = {1136073600, 1199145600, 1262304000, 1325376000, 1388534400, 1451606400, 1514764800},
xticklabels = {2006, 2008, 2010, 2012, 2014, 2016, 2018},
scaled ticks=false,
axis y line*=left,
axis x line*=bottom,
ytick = {0,1,2},
yticklabels = {0,1,2},
]
\addlegendimage{no markers, customG}
\addplot [thin, black!30, dotted]
table [row sep=\\]{%
	1089054280	1 \\
	1542654280	1 \\
};
\addplot [thick, customG]
table [row sep=\\]{%
1089054280	1 \\
1091646280	1 \\
1094238280	0 \\
1096830280	1 \\
1099422280	1 \\
1102014280	1.35714285714286 \\
1104606280	1.5 \\
1107198280	1.57142857142857 \\
1109790280	1.58333333333333 \\
1112382280	1.4375 \\
1114974280	1.21052631578947 \\
1117566280	1.31914893617021 \\
1120158280	1.28571428571429 \\
1122750280	1.23636363636364 \\
1125342280	1.14893617021277 \\
1127934280	1 \\
1130526280	0 \\
1133118280	0 \\
1135710280	0 \\
1138302280	1 \\
1140894280	2 \\
1143486280	2.07692307692308 \\
1146078280	2.1875 \\
1148670280	2.31578947368421 \\
1151262280	1.73913043478261 \\
1153854280	1.65217391304348 \\
1156446280	2.03703703703704 \\
1159038280	1.85185185185185 \\
1161630280	1.22222222222222 \\
1164222280	1.0625 \\
1166814280	1.24242424242424 \\
1169406280	1.25 \\
1171998280	1.64864864864865 \\
1174590280	1.76363636363636 \\
1177182280	1.65454545454545 \\
1179774280	1.22 \\
1182366280	0.921052631578947 \\
1184958280	1.1025641025641 \\
1187550280	1.14583333333333 \\
1190142280	1.0327868852459 \\
1192734280	0.971830985915493 \\
1195326280	0.894736842105263 \\
1197918280	0.757575757575758 \\
1200510280	0.875 \\
1203102280	0.65 \\
1205694280	0.660377358490566 \\
1208286280	0.743589743589744 \\
1210878280	1 \\
1213470280	1 \\
1216062280	0 \\
1218654280	0.0476190476190476 \\
1221246280	0.317073170731707 \\
1223838280	0.46875 \\
1226430280	0.578947368421053 \\
1229022280	0.71875 \\
1231614280	0.764705882352941 \\
1234206280	0.818181818181818 \\
1236798280	0.8 \\
1239390280	0.755555555555556 \\
1241982280	1.1063829787234 \\
1244574280	1.4375 \\
1247166280	1.234375 \\
1249758280	1.13698630136986 \\
1252350280	1.08 \\
1254942280	1.02739726027397 \\
1257534280	0.836065573770492 \\
1260126280	0.901639344262295 \\
1262718280	0.87719298245614 \\
1265310280	0.948275862068966 \\
1267902280	0.742424242424242 \\
1270494280	0.672131147540984 \\
1273086280	0.507462686567164 \\
1275678280	0.603174603174603 \\
1278270280	0.575757575757576 \\
1280862280	0.698412698412698 \\
1283454280	0.852459016393443 \\
1286046280	0.766233766233766 \\
1288638280	0.852941176470588 \\
1291230280	0.646341463414634 \\
1293822280	0.777777777777778 \\
1296414280	0.763888888888889 \\
1299006280	0.830985915492958 \\
1301598280	0.884057971014493 \\
1304190280	0.758064516129032 \\
1306782280	0.406779661016949 \\
1309374280	0.657534246575342 \\
1311966280	0.755555555555556 \\
1314558280	0.6875 \\
1317150280	0.65979381443299 \\
1319742280	0.75 \\
1322334280	0.833333333333333 \\
1324926280	0.978947368421053 \\
1327518280	0.802469135802469 \\
1330110280	0.896551724137931 \\
1332702280	0.985915492957746 \\
1335294280	0.917647058823529 \\
1337886280	0.663366336633663 \\
1340478280	0.663551401869159 \\
1343070280	0.653225806451613 \\
1345662280	0.676923076923077 \\
1348254280	0.602564102564103 \\
1350846280	0.626582278481013 \\
1353438280	0.576086956521739 \\
1356030280	0.446808510638298 \\
1358622280	0.446153846153846 \\
1361214280	0.539325842696629 \\
1363806280	0.744897959183674 \\
1366398280	0.704761904761905 \\
1368990280	0.621359223300971 \\
1371582280	0.409638554216867 \\
1374174280	0.413793103448276 \\
1376766280	0.494736842105263 \\
1379358280	0.679245283018868 \\
1381950280	0.7 \\
1384542280	0.510204081632653 \\
1387134280	0.554455445544555 \\
1389726280	0.54 \\
1392318280	0.442622950819672 \\
1394910280	0.454545454545455 \\
1397502280	0.428571428571429 \\
1400094280	0.406593406593407 \\
1402686280	0.387096774193548 \\
1405278280	0.402173913043478 \\
1407870280	0.48936170212766 \\
1410462280	0.53921568627451 \\
1413054280	0.404761904761905 \\
1415646280	0.394160583941606 \\
1418238280	0.409836065573771 \\
1420830280	0.504950495049505 \\
1423422280	0.48314606741573 \\
1426014280	0.56 \\
1428606280	0.424657534246575 \\
1431198280	0.446153846153846 \\
1433790280	0.377049180327869 \\
1436382280	0.32258064516129 \\
1438974280	0.218390804597701 \\
1441566280	0.26530612244898 \\
1444158280	0.4 \\
1446750280	0.432098765432099 \\
1449342280	0.337662337662338 \\
1451934280	0.4 \\
1454526280	0.346153846153846 \\
1457118280	0.519480519480519 \\
1459710280	0.45 \\
1462302280	0.368932038834951 \\
1464894280	0.321428571428571 \\
1467486280	0.266666666666667 \\
1470078280	0.246376811594203 \\
1472670280	0.255555555555556 \\
1475262280	0.361904761904762 \\
1477854280	0.438775510204082 \\
1480446280	0.4 \\
1483038280	0.508196721311475 \\
1485630280	0.597014925373134 \\
1488222280	0.384615384615385 \\
1490814280	0.364485981308411 \\
1493406280	0.373831775700935 \\
1495998280	0.382978723404255 \\
1498590280	0.293478260869565 \\
1501182280	0.25 \\
1503774280	0.325 \\
1506366280	0.291139240506329 \\
1508958280	0.373493975903614 \\
1511550280	0.394736842105263 \\
1514142280	0.39344262295082 \\
1516734280	0.134615384615385 \\
1519326280	0.320512820512821 \\
1521918280	0.40625 \\
1524510280	0.465346534653465 \\
1527102280	0.333333333333333 \\
1529694280	0.356164383561644 \\
1532286280	0.28 \\
1534878280	0.326530612244898 \\
1537470280	0.254545454545455 \\
1540062280	0.379746835443038 \\
1542654280	0.393258426966292 \\
};
\end{axis}
    	
    	\node at (-9.3,.15\textwidth) {\textbf{a}};
    	\node at (-1.6,.15\textwidth) {\textbf{b}};
    \end{tikzpicture}
    \caption{Comparative analysis of file-based co-authorship vs. line-based co-editing networks. \textbf{\sffamily a} Number of nodes and edges of networks aggregated over the entire project duration. Here, the co-authorship network overcounts relationships as editing the same file does not require a co-editing relationship on a line basis. \textbf{\sffamily b} Proportion of edges in both networks over a moving 90 day window. Here, the co-authorship network frequently does not display links present in the co-editing network, as with co-editing links interactions with developers not contributing code in the present time window can be considered.}
    \label{fig:network_type_edge_relation}
\end{figure*}
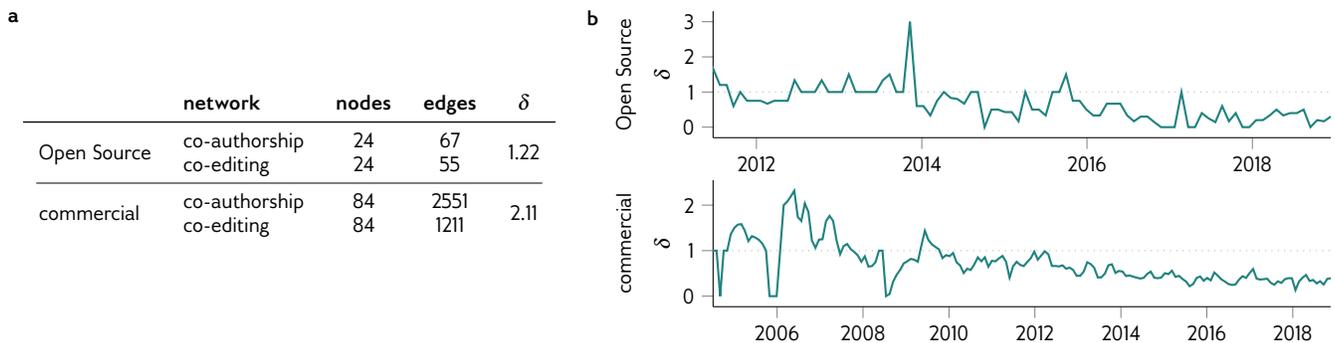

As outlined in section \ref{sec:related}, the analysis of co-authorship networks that capture which developers have contributed to the same files has received significant attention.
At the same time, recent works have argued for more fine-grained definitions of collaboration networks, using e.g. function points or code lines~\cite{Joblin2015,Scholtes2016}.
We contribute to this discussion and investigate the differences between a line- and a file-based approach to construct developer collaboration networks.
Our results show that (i) this choice of granularity has considerable influence on the resulting network topologies, (ii) that the resulting differences are project-dependent, and (iii) that the differences between the resulting networks exhibit temporal inhomogeneities.

For our analysis, we first use \texttt{git2net} to extract (a) a file-based co-editing network $G_f$ (which for simplicity we call co-authorship network), and (b) a line-based co-editing network $G_l$ for the Open Source project \texttt{igraph} as well as for a large commercial software project.
For both networks, we compare the time-aggregated projections (constructed as described in \ref{sec:results:networks}) and the sequence of networks obtained via a rolling window analysis.
For each time window (as well as for the time-aggregated network), we then quantitatively assess the difference between $G_l$ and $G_f$.
We first observe that the set of nodes in both networks is necessarily the same.
As a maximally simple approach to assess the difference between the two networks, we can thus calculate $\delta := \frac{m_f}{m_l}$, where $m_f$ and $m_l$ are the number of links in the file-based co-authorship network and the line-based co-editing networks, respectively.

Figure \ref{fig:network_type_edge_relation} shows the result of this analysis.
Fig. \ref{fig:network_type_edge_relation}a confirms that the file-based co-authorship network does not resolve where in the file edits take place, leading to a significantly higher number of links compared to the co-editing network in both projects.
We expect many of these additional links to be \emph{false positives}, in the sense that despite two developers having made edits to the same file no actual \emph{collaboration} on the same code actually occurred.

Fig. \ref{fig:network_type_edge_relation}b highlights the temporal dimension of these differences.
It shows the time-evolving difference between the two network abstractions, using a 90 day moving window.
For each window, the difference $\delta$ between the two networks is reported.
Importantly, we observe time windows where $\delta<1$, which indicates that the line-based co-editing networks feature additional links over the file-based co-authorship network.
This is due to the fact that a file-based (temporal) co-authorship network does not consider commits to files made outside the time window currently analysed.
However, our detailed analysis of co-edit relations can nevertheless identify that at time $t$ within the time window developer $A$ has edited code originally authored by developer $B$ in a commit outside the time window.
We argue that neglecting this relation introduces the risk of \emph{false negatives}, in the sense that we would omit the need of collaboration or coordination associated with a commit occurring at time $t$.
This subtle but important difference highlights the limitations of a simple file-based extraction of collaboration networks and showcases the advantage of our approach.

\subsection{Analysis of Temporal Co-Editing Networks}
\label{sec:results:temporal}
\newlength{\mywidth}
\newlength{\myheight}
\newlength{\myxdistance}
\newlength{\myydistance}
\setlength{\mywidth}{.51\textwidth}
\setlength{\myheight}{.17\textwidth}
\setlength{\myxdistance}{.49\textwidth}
\setlength{\myydistance}{-.10\textwidth}
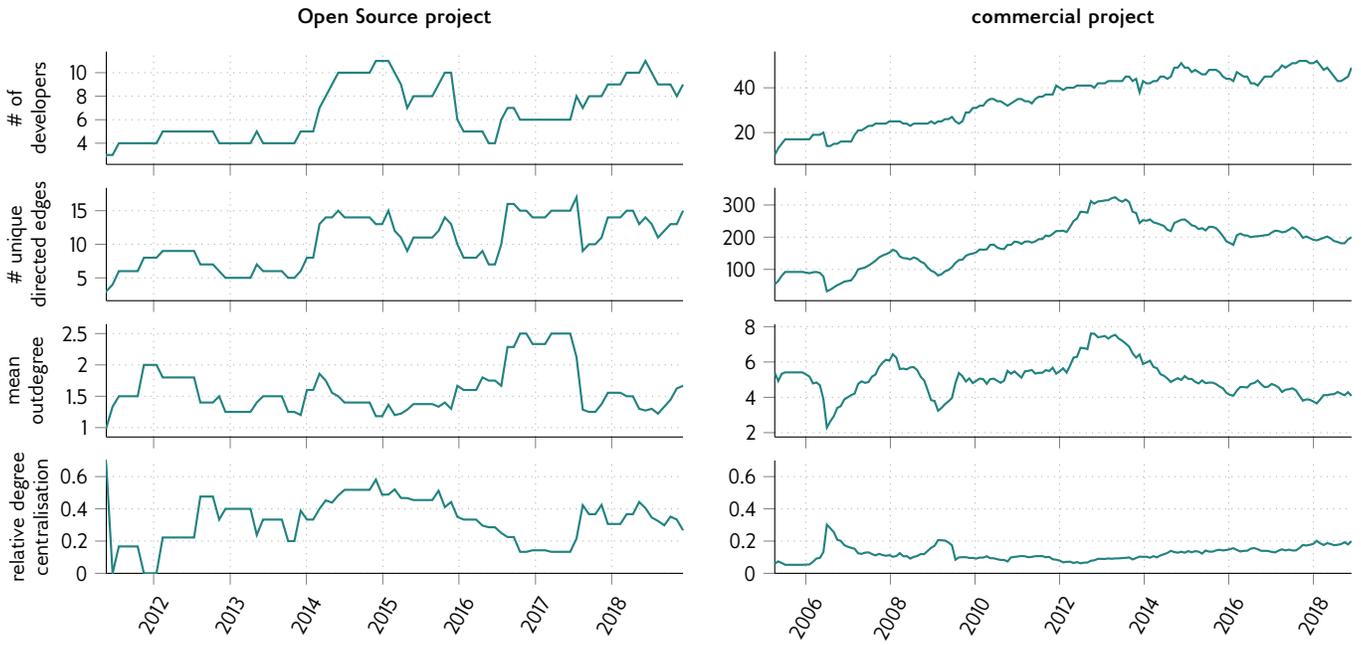
\begin{figure*}
	\centering
	\begin{tikzpicture}\footnotesize\sffamily
		\begin{axis}[
at={(0,0)},
width=\mywidth,
height=\myheight,
yticklabel style={font=\sansmath\sffamily},
ylabel style={align=center, yshift=3mm},
ylabel={\# of\\developers},
xmin=1305815836, xmax=1544279836,
tick align=outside,
tick pos=left,
ylabel style={align=center, yshift=-5mm},
x grid style={white!69.01960784313725!black},
y grid style={white!69.01960784313725!black},
axis y line*=left,
axis x line*=bottom,
xtick = {1136073600, 1199145600, 1262304000, 1325376000, 1357041600, 1388534400, 1420113600, 1451606400, 1483272000, 1514764800},
xticklabels = \empty,
scaled ticks=false,
title={\textbf{Open Source project}},
grid style={dotted,black!30},
grid
]
\addlegendimage{no markers, customG}
\addplot [thick, customG]
table [row sep=\\]{%
1160663836	2 \\
1163255836	2 \\
1165847836	2 \\
1168439836	2 \\
1171031836	2 \\
1173623836	2 \\
1176215836	2 \\
1178807836	2 \\
1181399836	2 \\
1183991836	2 \\
1186583836	2 \\
1189175836	2 \\
1191767836	2 \\
1194359836	2 \\
1196951836	2 \\
1199543836	2 \\
1202135836	2 \\
1204727836	2 \\
1207319836	2 \\
1209911836	2 \\
1212503836	2 \\
1215095836	2 \\
1217687836	2 \\
1220279836	2 \\
1222871836	2 \\
1225463836	2 \\
1228055836	2 \\
1230647836	2 \\
1233239836	2 \\
1235831836	2 \\
1238423836	2 \\
1241015836	2 \\
1243607836	2 \\
1246199836	2 \\
1248791836	2 \\
1251383836	2 \\
1253975836	2 \\
1256567836	2 \\
1259159836	2 \\
1261751836	2 \\
1264343836	2 \\
1266935836	2 \\
1269527836	2 \\
1272119836	2 \\
1274711836	2 \\
1277303836	2 \\
1279895836	2 \\
1282487836	2 \\
1285079836	2 \\
1287671836	2 \\
1290263836	2 \\
1292855836	2 \\
1295447836	2 \\
1298039836	2 \\
1300631836	2 \\
1303223836	2 \\
1305815836	3 \\
1308407836	3 \\
1310999836	4 \\
1313591836	4 \\
1316183836	4 \\
1318775836	4 \\
1321367836	4 \\
1323959836	4 \\
1326551836	4 \\
1329143836	5 \\
1331735836	5 \\
1334327836	5 \\
1336919836	5 \\
1339511836	5 \\
1342103836	5 \\
1344695836	5 \\
1347287836	5 \\
1349879836	5 \\
1352471836	4 \\
1355063836	4 \\
1357655836	4 \\
1360247836	4 \\
1362839836	4 \\
1365431836	4 \\
1368023836	5 \\
1370615836	4 \\
1373207836	4 \\
1375799836	4 \\
1378391836	4 \\
1380983836	4 \\
1383575836	4 \\
1386167836	5 \\
1388759836	5 \\
1391351836	5 \\
1393943836	7 \\
1396535836	8 \\
1399127836	9 \\
1401719836	10 \\
1404311836	10 \\
1406903836	10 \\
1409495836	10 \\
1412087836	10 \\
1414679836	10 \\
1417271836	11 \\
1419863836	11 \\
1422455836	11 \\
1425047836	10 \\
1427639836	9 \\
1430231836	7 \\
1432823836	8 \\
1435415836	8 \\
1438007836	8 \\
1440599836	8 \\
1443191836	9 \\
1445783836	10 \\
1448375836	10 \\
1450967836	6 \\
1453559836	5 \\
1456151836	5 \\
1458743836	5 \\
1461335836	5 \\
1463927836	4 \\
1466519836	4 \\
1469111836	6 \\
1471703836	7 \\
1474295836	7 \\
1476887836	6 \\
1479479836	6 \\
1482071836	6 \\
1484663836	6 \\
1487255836	6 \\
1489847836	6 \\
1492439836	6 \\
1495031836	6 \\
1497623836	6 \\
1500215836	8 \\
1502807836	7 \\
1505399836	8 \\
1507991836	8 \\
1510583836	8 \\
1513175836	9 \\
1515767836	9 \\
1518359836	9 \\
1520951836	10 \\
1523543836	10 \\
1526135836	10 \\
1528727836	11 \\
1531319836	10 \\
1533911836	9 \\
1536503836	9 \\
1539095836	9 \\
1541687836	8 \\
1544279836	9 \\
};
\end{axis}
		\begin{axis}[
at={(0,\myydistance)},
width=\mywidth,
height=\myheight,
yticklabel style={font=\sansmath\sffamily},
ylabel style={align=center, yshift=3mm},
ylabel={\# unique\\directed edges},
xmin=1305815836, xmax=1544279836,
tick align=outside,
tick pos=left,
ylabel style={align=center, yshift=-5mm},
x grid style={white!69.01960784313725!black},
y grid style={white!69.01960784313725!black},
axis y line*=left,
axis x line*=bottom,
xtick = {1136073600, 1199145600, 1262304000, 1325376000, 1357041600, 1388534400, 1420113600, 1451606400, 1483272000, 1514764800},
xticklabels = \empty,
xticklabel style={rotate=60},
scaled ticks=false,
grid style={dotted,black!30},
grid
]
\addlegendimage{no markers, customG}
\addplot [thick, customG]
table [row sep=\\]{%
1160663836	2 \\
1163255836	2 \\
1165847836	2 \\
1168439836	2 \\
1171031836	2 \\
1173623836	2 \\
1176215836	2 \\
1178807836	2 \\
1181399836	2 \\
1183991836	2 \\
1186583836	2 \\
1189175836	2 \\
1191767836	2 \\
1194359836	2 \\
1196951836	2 \\
1199543836	2 \\
1202135836	2 \\
1204727836	2 \\
1207319836	2 \\
1209911836	2 \\
1212503836	2 \\
1215095836	2 \\
1217687836	2 \\
1220279836	2 \\
1222871836	2 \\
1225463836	2 \\
1228055836	2 \\
1230647836	2 \\
1233239836	2 \\
1235831836	2 \\
1238423836	2 \\
1241015836	2 \\
1243607836	2 \\
1246199836	2 \\
1248791836	2 \\
1251383836	2 \\
1253975836	2 \\
1256567836	2 \\
1259159836	2 \\
1261751836	2 \\
1264343836	2 \\
1266935836	2 \\
1269527836	2 \\
1272119836	2 \\
1274711836	2 \\
1277303836	2 \\
1279895836	2 \\
1282487836	2 \\
1285079836	2 \\
1287671836	2 \\
1290263836	2 \\
1292855836	2 \\
1295447836	2 \\
1298039836	2 \\
1300631836	2 \\
1303223836	2 \\
1305815836	3 \\
1308407836	4 \\
1310999836	6 \\
1313591836	6 \\
1316183836	6 \\
1318775836	6 \\
1321367836	8 \\
1323959836	8 \\
1326551836	8 \\
1329143836	9 \\
1331735836	9 \\
1334327836	9 \\
1336919836	9 \\
1339511836	9 \\
1342103836	9 \\
1344695836	7 \\
1347287836	7 \\
1349879836	7 \\
1352471836	6 \\
1355063836	5 \\
1357655836	5 \\
1360247836	5 \\
1362839836	5 \\
1365431836	5 \\
1368023836	7 \\
1370615836	6 \\
1373207836	6 \\
1375799836	6 \\
1378391836	6 \\
1380983836	5 \\
1383575836	5 \\
1386167836	6 \\
1388759836	8 \\
1391351836	8 \\
1393943836	13 \\
1396535836	14 \\
1399127836	14 \\
1401719836	15 \\
1404311836	14 \\
1406903836	14 \\
1409495836	14 \\
1412087836	14 \\
1414679836	14 \\
1417271836	13 \\
1419863836	13 \\
1422455836	15 \\
1425047836	12 \\
1427639836	11 \\
1430231836	9 \\
1432823836	11 \\
1435415836	11 \\
1438007836	11 \\
1440599836	11 \\
1443191836	12 \\
1445783836	14 \\
1448375836	13 \\
1450967836	10 \\
1453559836	8 \\
1456151836	8 \\
1458743836	8 \\
1461335836	9 \\
1463927836	7 \\
1466519836	7 \\
1469111836	10 \\
1471703836	16 \\
1474295836	16 \\
1476887836	15 \\
1479479836	15 \\
1482071836	14 \\
1484663836	14 \\
1487255836	14 \\
1489847836	15 \\
1492439836	15 \\
1495031836	15 \\
1497623836	15 \\
1500215836	17 \\
1502807836	9 \\
1505399836	10 \\
1507991836	10 \\
1510583836	11 \\
1513175836	14 \\
1515767836	14 \\
1518359836	14 \\
1520951836	15 \\
1523543836	15 \\
1526135836	13 \\
1528727836	14 \\
1531319836	13 \\
1533911836	11 \\
1536503836	12 \\
1539095836	13 \\
1541687836	13 \\
1544279836	15 \\
};
\end{axis}
		\begin{axis}[
at={(0,2*\myydistance)},
width=\mywidth,
height=\myheight,
yticklabel style={font=\sansmath\sffamily},
ylabel style={align=center, yshift=3mm},
ylabel={mean\\outdegree},
xmin=1305815836, xmax=1544279836,
tick align=outside,
tick pos=left,
ylabel style={align=center, yshift=-5mm},
x grid style={white!69.01960784313725!black},
y grid style={white!69.01960784313725!black},
axis y line*=left,
axis x line*=bottom,
xtick = {1136073600, 1199145600, 1262304000, 1325376000, 1357041600, 1388534400, 1420113600, 1451606400, 1483272000, 1514764800},
xticklabels = \empty,
xticklabel style={rotate=60},
scaled ticks=false,
grid style={dotted,black!30},
grid
]
\addlegendimage{no markers, customG}
\addplot [thick, customG]
table [row sep=\\]{%
1160663836	1 \\
1163255836	1 \\
1165847836	1 \\
1168439836	1 \\
1171031836	1 \\
1173623836	1 \\
1176215836	1 \\
1178807836	1 \\
1181399836	1 \\
1183991836	1 \\
1186583836	1 \\
1189175836	1 \\
1191767836	1 \\
1194359836	1 \\
1196951836	1 \\
1199543836	1 \\
1202135836	1 \\
1204727836	1 \\
1207319836	1 \\
1209911836	1 \\
1212503836	1 \\
1215095836	1 \\
1217687836	1 \\
1220279836	1 \\
1222871836	1 \\
1225463836	1 \\
1228055836	1 \\
1230647836	1 \\
1233239836	1 \\
1235831836	1 \\
1238423836	1 \\
1241015836	1 \\
1243607836	1 \\
1246199836	1 \\
1248791836	1 \\
1251383836	1 \\
1253975836	1 \\
1256567836	1 \\
1259159836	1 \\
1261751836	1 \\
1264343836	1 \\
1266935836	1 \\
1269527836	1 \\
1272119836	1 \\
1274711836	1 \\
1277303836	1 \\
1279895836	1 \\
1282487836	1 \\
1285079836	1 \\
1287671836	1 \\
1290263836	1 \\
1292855836	1 \\
1295447836	1 \\
1298039836	1 \\
1300631836	1 \\
1303223836	1 \\
1305815836	1 \\
1308407836	1.33333333333333 \\
1310999836	1.5 \\
1313591836	1.5 \\
1316183836	1.5 \\
1318775836	1.5 \\
1321367836	2 \\
1323959836	2 \\
1326551836	2 \\
1329143836	1.8 \\
1331735836	1.8 \\
1334327836	1.8 \\
1336919836	1.8 \\
1339511836	1.8 \\
1342103836	1.8 \\
1344695836	1.4 \\
1347287836	1.4 \\
1349879836	1.4 \\
1352471836	1.5 \\
1355063836	1.25 \\
1357655836	1.25 \\
1360247836	1.25 \\
1362839836	1.25 \\
1365431836	1.25 \\
1368023836	1.4 \\
1370615836	1.5 \\
1373207836	1.5 \\
1375799836	1.5 \\
1378391836	1.5 \\
1380983836	1.25 \\
1383575836	1.25 \\
1386167836	1.2 \\
1388759836	1.6 \\
1391351836	1.6 \\
1393943836	1.85714285714286 \\
1396535836	1.75 \\
1399127836	1.55555555555556 \\
1401719836	1.5 \\
1404311836	1.4 \\
1406903836	1.4 \\
1409495836	1.4 \\
1412087836	1.4 \\
1414679836	1.4 \\
1417271836	1.18181818181818 \\
1419863836	1.18181818181818 \\
1422455836	1.36363636363636 \\
1425047836	1.2 \\
1427639836	1.22222222222222 \\
1430231836	1.28571428571429 \\
1432823836	1.375 \\
1435415836	1.375 \\
1438007836	1.375 \\
1440599836	1.375 \\
1443191836	1.33333333333333 \\
1445783836	1.4 \\
1448375836	1.3 \\
1450967836	1.66666666666667 \\
1453559836	1.6 \\
1456151836	1.6 \\
1458743836	1.6 \\
1461335836	1.8 \\
1463927836	1.75 \\
1466519836	1.75 \\
1469111836	1.66666666666667 \\
1471703836	2.28571428571429 \\
1474295836	2.28571428571429 \\
1476887836	2.5 \\
1479479836	2.5 \\
1482071836	2.33333333333333 \\
1484663836	2.33333333333333 \\
1487255836	2.33333333333333 \\
1489847836	2.5 \\
1492439836	2.5 \\
1495031836	2.5 \\
1497623836	2.5 \\
1500215836	2.125 \\
1502807836	1.28571428571429 \\
1505399836	1.25 \\
1507991836	1.25 \\
1510583836	1.375 \\
1513175836	1.55555555555556 \\
1515767836	1.55555555555556 \\
1518359836	1.55555555555556 \\
1520951836	1.5 \\
1523543836	1.5 \\
1526135836	1.3 \\
1528727836	1.27272727272727 \\
1531319836	1.3 \\
1533911836	1.22222222222222 \\
1536503836	1.33333333333333 \\
1539095836	1.44444444444444 \\
1541687836	1.625 \\
1544279836	1.66666666666667 \\
};
\end{axis}
		\begin{axis}[
at={(0,3*\myydistance)},
width=\mywidth,
height=\myheight,
yticklabel style={font=\sansmath\sffamily},
ylabel style={align=center, yshift=3mm},
ylabel={relative degree\\centralisation},
xmin=1305815836, xmax=1544279836,
ymin=0, ymax=.7,
tick align=outside,
tick pos=left,
ylabel style={align=center, yshift=-5mm},
x grid style={white!69.01960784313725!black},
y grid style={white!69.01960784313725!black},
axis y line*=left,
axis x line*=bottom,
xtick = {1136073600, 1199145600, 1262304000, 1325376000, 1357041600, 1388534400, 1420113600, 1451606400, 1483272000, 1514764800},
xticklabels = {2006, 2008, 2010, 2012, 2013, 2014, 2015, 2016, 2017, 2018},
xticklabel style={rotate=60},
scaled ticks=false,
grid style={dotted,black!30},
grid
]
\addlegendimage{no markers, customG}
\addplot [thick, customG]
table [row sep=\\]{%
1160663836	1 \\
1163255836	1 \\
1165847836	1 \\
1168439836	1 \\
1171031836	1 \\
1173623836	1 \\
1176215836	1 \\
1178807836	1 \\
1181399836	1 \\
1183991836	1 \\
1186583836	1 \\
1189175836	1 \\
1191767836	1 \\
1194359836	1 \\
1196951836	1 \\
1199543836	1 \\
1202135836	1 \\
1204727836	1 \\
1207319836	1 \\
1209911836	1 \\
1212503836	1 \\
1215095836	1 \\
1217687836	1 \\
1220279836	1 \\
1222871836	1 \\
1225463836	1 \\
1228055836	1 \\
1230647836	1 \\
1233239836	1 \\
1235831836	1 \\
1238423836	1 \\
1241015836	1 \\
1243607836	1 \\
1246199836	1 \\
1248791836	1 \\
1251383836	1 \\
1253975836	1 \\
1256567836	1 \\
1259159836	1 \\
1261751836	1 \\
1264343836	1 \\
1266935836	1 \\
1269527836	1 \\
1272119836	1 \\
1274711836	1 \\
1277303836	1 \\
1279895836	1 \\
1282487836	1 \\
1285079836	1 \\
1287671836	1 \\
1290263836	1 \\
1292855836	1 \\
1295447836	1 \\
1298039836	1 \\
1300631836	1 \\
1303223836	1 \\
1305815836	0.666666666666667 \\
1308407836	0 \\
1310999836	0.166666666666667 \\
1313591836	0.166666666666667 \\
1316183836	0.166666666666667 \\
1318775836	0.166666666666667 \\
1321367836	0 \\
1323959836	0 \\
1326551836	0 \\
1329143836	0.222222222222222 \\
1331735836	0.222222222222222 \\
1334327836	0.222222222222222 \\
1336919836	0.222222222222222 \\
1339511836	0.222222222222222 \\
1342103836	0.222222222222222 \\
1344695836	0.476190476190476 \\
1347287836	0.476190476190476 \\
1349879836	0.476190476190476 \\
1352471836	0.333333333333333 \\
1355063836	0.4 \\
1357655836	0.4 \\
1360247836	0.4 \\
1362839836	0.4 \\
1365431836	0.4 \\
1368023836	0.238095238095238 \\
1370615836	0.333333333333333 \\
1373207836	0.333333333333333 \\
1375799836	0.333333333333333 \\
1378391836	0.333333333333333 \\
1380983836	0.2 \\
1383575836	0.2 \\
1386167836	0.388888888888889 \\
1388759836	0.333333333333333 \\
1391351836	0.333333333333333 \\
1393943836	0.4 \\
1396535836	0.452380952380952 \\
1399127836	0.438775510204082 \\
1401719836	0.483333333333333 \\
1404311836	0.517857142857143 \\
1406903836	0.517857142857143 \\
1409495836	0.517857142857143 \\
1412087836	0.517857142857143 \\
1414679836	0.517857142857143 \\
1417271836	0.581196581196581 \\
1419863836	0.487179487179487 \\
1422455836	0.488888888888889 \\
1425047836	0.520833333333333 \\
1427639836	0.467532467532468 \\
1430231836	0.466666666666667 \\
1432823836	0.454545454545455 \\
1435415836	0.454545454545455 \\
1438007836	0.454545454545455 \\
1440599836	0.454545454545455 \\
1443191836	0.511904761904762 \\
1445783836	0.410714285714286 \\
1448375836	0.442307692307692 \\
1450967836	0.35 \\
1453559836	0.333333333333333 \\
1456151836	0.333333333333333 \\
1458743836	0.333333333333333 \\
1461335836	0.296296296296296 \\
1463927836	0.285714285714286 \\
1466519836	0.285714285714286 \\
1469111836	0.25 \\
1471703836	0.225 \\
1474295836	0.225 \\
1476887836	0.133333333333333 \\
1479479836	0.133333333333333 \\
1482071836	0.142857142857143 \\
1484663836	0.142857142857143 \\
1487255836	0.142857142857143 \\
1489847836	0.133333333333333 \\
1492439836	0.133333333333333 \\
1495031836	0.133333333333333 \\
1497623836	0.133333333333333 \\
1500215836	0.215686274509804 \\
1502807836	0.422222222222222 \\
1505399836	0.366666666666667 \\
1507991836	0.366666666666667 \\
1510583836	0.424242424242424 \\
1513175836	0.306122448979592 \\
1515767836	0.306122448979592 \\
1518359836	0.306122448979592 \\
1520951836	0.366666666666667 \\
1523543836	0.366666666666667 \\
1526135836	0.442307692307692 \\
1528727836	0.404761904761905 \\
1531319836	0.346153846153846 \\
1533911836	0.324675324675325 \\
1536503836	0.297619047619048 \\
1539095836	0.351648351648352 \\
1541687836	0.333333333333333 \\
1544279836	0.266666666666667 \\
};
\end{axis}
		\begin{axis}[
at={(\myxdistance,0)},
width=\mywidth,
height=\myheight,
axis y line*=left,
ylabel near ticks,
yticklabel style={font=\sansmath\sffamily},
xmin=1112896216, xmax=1543168216,
tick align=outside,
xtick = {1136073600, 1199145600, 1262304000, 1325376000, 1388534400, 1451606400, 1514764800},
xticklabels = \empty, 
xticklabel style={rotate=60},
scaled ticks=false,
axis x line*=bottom,
title={\textbf{commercial project}},
grid style={dotted,black!30},
grid
]
\addlegendimage{no markers, customG}
\addplot [thick, customG]
table [row sep=\\]{%
1112896216	10 \\
1115488216	13 \\
1118080216	15 \\
1120672216	17 \\
1123264216	17 \\
1125856216	17 \\
1128448216	17 \\
1131040216	17 \\
1133632216	17 \\
1136224216	17 \\
1138816216	17 \\
1141408216	19 \\
1144000216	19 \\
1146592216	19 \\
1149184216	20 \\
1151776216	14 \\
1154368216	14 \\
1156960216	15 \\
1159552216	15 \\
1162144216	16 \\
1164736216	16 \\
1167328216	16 \\
1169920216	16 \\
1172512216	19 \\
1175104216	21 \\
1177696216	21 \\
1180288216	22 \\
1182880216	23 \\
1185472216	23 \\
1188064216	24 \\
1190656216	24 \\
1193248216	24 \\
1195840216	24 \\
1198432216	25 \\
1201024216	25 \\
1203616216	25 \\
1206208216	25 \\
1208800216	24 \\
1211392216	24 \\
1213984216	23 \\
1216576216	24 \\
1219168216	24 \\
1221760216	24 \\
1224352216	24 \\
1226944216	24 \\
1229536216	25 \\
1232128216	24 \\
1234720216	25 \\
1237312216	25 \\
1239904216	26 \\
1242496216	26 \\
1245088216	27 \\
1247680216	25 \\
1250272216	24 \\
1252864216	25 \\
1255456216	29 \\
1258048216	29 \\
1260640216	31 \\
1263232216	31 \\
1265824216	32 \\
1268416216	32 \\
1271008216	34 \\
1273600216	35 \\
1276192216	35 \\
1278784216	34 \\
1281376216	34 \\
1283968216	33 \\
1286560216	32 \\
1289152216	33 \\
1291744216	34 \\
1294336216	35 \\
1296928216	35 \\
1299520216	34 \\
1302112216	34 \\
1304704216	33 \\
1307296216	35 \\
1309888216	36 \\
1312480216	36 \\
1315072216	37 \\
1317664216	37 \\
1320256216	37 \\
1322848216	41 \\
1325440216	40 \\
1328032216	39 \\
1330624216	40 \\
1333216216	40 \\
1335808216	40 \\
1338400216	41 \\
1340992216	41 \\
1343584216	41 \\
1346176216	41 \\
1348768216	41 \\
1351360216	40 \\
1353952216	42 \\
1356544216	42 \\
1359136216	42 \\
1361728216	43 \\
1364320216	43 \\
1366912216	43 \\
1369504216	43 \\
1372096216	43 \\
1374688216	45 \\
1377280216	45 \\
1379872216	43 \\
1382464216	44 \\
1385056216	38 \\
1387648216	43 \\
1390240216	42 \\
1392832216	42 \\
1395424216	43 \\
1398016216	43 \\
1400608216	45 \\
1403200216	45 \\
1405792216	44 \\
1408384216	45 \\
1410976216	49 \\
1413568216	49 \\
1416160216	51 \\
1418752216	49 \\
1421344216	49 \\
1423936216	47 \\
1426528216	48 \\
1429120216	47 \\
1431712216	46 \\
1434304216	46 \\
1436896216	48 \\
1439488216	48 \\
1442080216	48 \\
1444672216	47 \\
1447264216	45 \\
1449856216	44 \\
1452448216	44 \\
1455040216	43 \\
1457632216	47 \\
1460224216	46 \\
1462816216	45 \\
1465408216	45 \\
1468000216	42 \\
1470592216	42 \\
1473184216	41 \\
1475776216	43 \\
1478368216	45 \\
1480960216	45 \\
1483552216	45 \\
1486144216	47 \\
1488736216	48 \\
1491328216	50 \\
1493920216	49 \\
1496512216	50 \\
1499104216	51 \\
1501696216	51 \\
1504288216	52 \\
1506880216	52 \\
1509472216	52 \\
1512064216	51 \\
1514656216	51 \\
1517248216	52 \\
1519840216	50 \\
1522432216	48 \\
1525024216	49 \\
1527616216	47 \\
1530208216	45 \\
1532800216	43 \\
1535392216	43 \\
1537984216	44 \\
1540576216	45 \\
1543168216	49 \\
};
\end{axis}
		\begin{axis}[
at={(\myxdistance,\myydistance)},
width=\mywidth,
height=\myheight,
yticklabel style={font=\sansmath\sffamily},
xmin=1112896216, xmax=1543168216,
tick align=outside,
ylabel style={align=center, yshift=-5mm},
xtick = {1136073600, 1199145600, 1262304000, 1325376000, 1388534400, 1451606400, 1514764800},
xticklabels = \empty, 
xticklabel style={rotate=60},
scaled ticks=false,
axis y line*=left,
axis x line*=bottom,
grid style={dotted,black!30},
grid
]
\addlegendimage{no markers, customG}
\addplot [thick, customG]
table [row sep=\\]{%
1112896216	54 \\
1115488216	64 \\
1118080216	80 \\
1120672216	92 \\
1123264216	92 \\
1125856216	92 \\
1128448216	92 \\
1131040216	92 \\
1133632216	92 \\
1136224216	90 \\
1138816216	88 \\
1141408216	91 \\
1144000216	92 \\
1146592216	89 \\
1149184216	78 \\
1151776216	32 \\
1154368216	37 \\
1156960216	44 \\
1159552216	51 \\
1162144216	56 \\
1164736216	62 \\
1167328216	64 \\
1169920216	66 \\
1172512216	80 \\
1175104216	100 \\
1177696216	103 \\
1180288216	106 \\
1182880216	112 \\
1185472216	119 \\
1188064216	127 \\
1190656216	137 \\
1193248216	143 \\
1195840216	147 \\
1198432216	152 \\
1201024216	161 \\
1203616216	156 \\
1206208216	140 \\
1208800216	135 \\
1211392216	134 \\
1213984216	131 \\
1216576216	137 \\
1219168216	133 \\
1221760216	124 \\
1224352216	119 \\
1226944216	106 \\
1229536216	96 \\
1232128216	91 \\
1234720216	81 \\
1237312216	85 \\
1239904216	94 \\
1242496216	98 \\
1245088216	107 \\
1247680216	120 \\
1250272216	129 \\
1252864216	130 \\
1255456216	142 \\
1258048216	147 \\
1260640216	149 \\
1263232216	153 \\
1265824216	162 \\
1268416216	161 \\
1271008216	162 \\
1273600216	176 \\
1276192216	177 \\
1278784216	168 \\
1281376216	164 \\
1283968216	163 \\
1286560216	176 \\
1289152216	176 \\
1291744216	186 \\
1294336216	185 \\
1296928216	179 \\
1299520216	186 \\
1302112216	187 \\
1304704216	183 \\
1307296216	187 \\
1309888216	194 \\
1312480216	194 \\
1315072216	205 \\
1317664216	203 \\
1320256216	210 \\
1322848216	219 \\
1325440216	219 \\
1328032216	220 \\
1330624216	216 \\
1333216216	232 \\
1335808216	250 \\
1338400216	258 \\
1340992216	279 \\
1343584216	278 \\
1346176216	276 \\
1348768216	312 \\
1351360216	304 \\
1353952216	311 \\
1356544216	312 \\
1359136216	314 \\
1361728216	315 \\
1364320216	321 \\
1366912216	324 \\
1369504216	315 \\
1372096216	310 \\
1374688216	317 \\
1377280216	309 \\
1379872216	279 \\
1382464216	275 \\
1385056216	244 \\
1387648216	253 \\
1390240216	251 \\
1392832216	255 \\
1395424216	246 \\
1398016216	243 \\
1400608216	239 \\
1403200216	235 \\
1405792216	223 \\
1408384216	219 \\
1410976216	244 \\
1413568216	249 \\
1416160216	254 \\
1418752216	255 \\
1421344216	247 \\
1423936216	237 \\
1426528216	234 \\
1429120216	225 \\
1431712216	228 \\
1434304216	221 \\
1436896216	232 \\
1439488216	232 \\
1442080216	228 \\
1444672216	217 \\
1447264216	203 \\
1449856216	187 \\
1452448216	182 \\
1455040216	176 \\
1457632216	206 \\
1460224216	211 \\
1462816216	206 \\
1465408216	205 \\
1468000216	200 \\
1470592216	202 \\
1473184216	203 \\
1475776216	204 \\
1478368216	206 \\
1480960216	207 \\
1483552216	214 \\
1486144216	220 \\
1488736216	219 \\
1491328216	215 \\
1493920216	217 \\
1496512216	223 \\
1499104216	230 \\
1501696216	224 \\
1504288216	213 \\
1506880216	198 \\
1509472216	202 \\
1512064216	197 \\
1514656216	192 \\
1517248216	190 \\
1519840216	194 \\
1522432216	198 \\
1525024216	202 \\
1527616216	196 \\
1530208216	188 \\
1532800216	185 \\
1535392216	181 \\
1537984216	181 \\
1540576216	193 \\
1543168216	200 \\
};
\end{axis}
		\begin{axis}[
at={(\myxdistance,2*\myydistance)},
width=\mywidth,
height=\myheight,
yticklabel style={font=\sansmath\sffamily},
xmin=1112896216, xmax=1543168216,
tick align=outside,
ylabel style={align=center, yshift=-5mm},
xtick = {1136073600, 1199145600, 1262304000, 1325376000, 1388534400, 1451606400, 1514764800},
xticklabels = \empty, 
xticklabel style={rotate=60},
scaled ticks=false,
axis y line*=left,
axis x line*=bottom,
grid style={dotted,black!30},
grid
]
\addlegendimage{no markers, customG}
\addplot [thick, customG]
table [row sep=\\]{%
1112896216	5.4 \\
1115488216	4.92307692307692 \\
1118080216	5.33333333333333 \\
1120672216	5.41176470588235 \\
1123264216	5.41176470588235 \\
1125856216	5.41176470588235 \\
1128448216	5.41176470588235 \\
1131040216	5.41176470588235 \\
1133632216	5.41176470588235 \\
1136224216	5.29411764705882 \\
1138816216	5.17647058823529 \\
1141408216	4.78947368421053 \\
1144000216	4.84210526315789 \\
1146592216	4.68421052631579 \\
1149184216	3.9 \\
1151776216	2.28571428571429 \\
1154368216	2.64285714285714 \\
1156960216	2.93333333333333 \\
1159552216	3.4 \\
1162144216	3.5 \\
1164736216	3.875 \\
1167328216	4 \\
1169920216	4.125 \\
1172512216	4.21052631578947 \\
1175104216	4.76190476190476 \\
1177696216	4.90476190476191 \\
1180288216	4.81818181818182 \\
1182880216	4.8695652173913 \\
1185472216	5.17391304347826 \\
1188064216	5.29166666666667 \\
1190656216	5.70833333333333 \\
1193248216	5.95833333333333 \\
1195840216	6.125 \\
1198432216	6.08 \\
1201024216	6.44 \\
1203616216	6.24 \\
1206208216	5.6 \\
1208800216	5.625 \\
1211392216	5.58333333333333 \\
1213984216	5.69565217391304 \\
1216576216	5.70833333333333 \\
1219168216	5.54166666666667 \\
1221760216	5.16666666666667 \\
1224352216	4.95833333333333 \\
1226944216	4.41666666666667 \\
1229536216	3.84 \\
1232128216	3.79166666666667 \\
1234720216	3.24 \\
1237312216	3.4 \\
1239904216	3.61538461538462 \\
1242496216	3.76923076923077 \\
1245088216	3.96296296296296 \\
1247680216	4.8 \\
1250272216	5.375 \\
1252864216	5.2 \\
1255456216	4.89655172413793 \\
1258048216	5.06896551724138 \\
1260640216	4.80645161290323 \\
1263232216	4.93548387096774 \\
1265824216	5.0625 \\
1268416216	5.03125 \\
1271008216	4.76470588235294 \\
1273600216	5.02857142857143 \\
1276192216	5.05714285714286 \\
1278784216	4.94117647058824 \\
1281376216	4.82352941176471 \\
1283968216	4.93939393939394 \\
1286560216	5.5 \\
1289152216	5.33333333333333 \\
1291744216	5.47058823529412 \\
1294336216	5.28571428571429 \\
1296928216	5.11428571428571 \\
1299520216	5.47058823529412 \\
1302112216	5.5 \\
1304704216	5.54545454545455 \\
1307296216	5.34285714285714 \\
1309888216	5.38888888888889 \\
1312480216	5.38888888888889 \\
1315072216	5.54054054054054 \\
1317664216	5.48648648648649 \\
1320256216	5.67567567567568 \\
1322848216	5.34146341463415 \\
1325440216	5.475 \\
1328032216	5.64102564102564 \\
1330624216	5.4 \\
1333216216	5.8 \\
1335808216	6.25 \\
1338400216	6.29268292682927 \\
1340992216	6.80487804878049 \\
1343584216	6.78048780487805 \\
1346176216	6.73170731707317 \\
1348768216	7.60975609756098 \\
1351360216	7.6 \\
1353952216	7.40476190476191 \\
1356544216	7.42857142857143 \\
1359136216	7.47619047619048 \\
1361728216	7.32558139534884 \\
1364320216	7.46511627906977 \\
1366912216	7.53488372093023 \\
1369504216	7.32558139534884 \\
1372096216	7.2093023255814 \\
1374688216	7.04444444444444 \\
1377280216	6.86666666666667 \\
1379872216	6.48837209302326 \\
1382464216	6.25 \\
1385056216	6.42105263157895 \\
1387648216	5.88372093023256 \\
1390240216	5.97619047619048 \\
1392832216	6.07142857142857 \\
1395424216	5.72093023255814 \\
1398016216	5.65116279069767 \\
1400608216	5.31111111111111 \\
1403200216	5.22222222222222 \\
1405792216	5.06818181818182 \\
1408384216	4.86666666666667 \\
1410976216	4.97959183673469 \\
1413568216	5.08163265306122 \\
1416160216	4.98039215686275 \\
1418752216	5.20408163265306 \\
1421344216	5.04081632653061 \\
1423936216	5.04255319148936 \\
1426528216	4.875 \\
1429120216	4.78723404255319 \\
1431712216	4.95652173913043 \\
1434304216	4.80434782608696 \\
1436896216	4.83333333333333 \\
1439488216	4.83333333333333 \\
1442080216	4.75 \\
1444672216	4.61702127659574 \\
1447264216	4.51111111111111 \\
1449856216	4.25 \\
1452448216	4.13636363636364 \\
1455040216	4.09302325581395 \\
1457632216	4.38297872340426 \\
1460224216	4.58695652173913 \\
1462816216	4.57777777777778 \\
1465408216	4.55555555555556 \\
1468000216	4.76190476190476 \\
1470592216	4.80952380952381 \\
1473184216	4.95121951219512 \\
1475776216	4.74418604651163 \\
1478368216	4.57777777777778 \\
1480960216	4.6 \\
1483552216	4.75555555555556 \\
1486144216	4.68085106382979 \\
1488736216	4.5625 \\
1491328216	4.3 \\
1493920216	4.42857142857143 \\
1496512216	4.46 \\
1499104216	4.50980392156863 \\
1501696216	4.3921568627451 \\
1504288216	4.09615384615385 \\
1506880216	3.80769230769231 \\
1509472216	3.88461538461538 \\
1512064216	3.86274509803922 \\
1514656216	3.76470588235294 \\
1517248216	3.65384615384615 \\
1519840216	3.88 \\
1522432216	4.125 \\
1525024216	4.12244897959184 \\
1527616216	4.17021276595745 \\
1530208216	4.17777777777778 \\
1532800216	4.30232558139535 \\
1535392216	4.2093023255814 \\
1537984216	4.11363636363636 \\
1540576216	4.28888888888889 \\
1543168216	4.08163265306122 \\
};
\end{axis}
		\begin{axis}[
at={(\myxdistance,3*\myydistance)},
width=\mywidth,
height=\myheight,
ymin=0, ymax=.7,
yticklabel style={font=\sansmath\sffamily},
xmin=1112896216, xmax=1543168216,
tick align=outside,
ylabel style={align=center, yshift=-5mm},
xtick = {1136073600, 1199145600, 1262304000, 1325376000, 1388534400, 1451606400, 1514764800},
xticklabels = {2006, 2008, 2010, 2012, 2014, 2016, 2018},
xticklabel style={rotate=60},
scaled ticks=false,
axis y line*=left,
axis x line*=bottom,
grid style={dotted,black!30},
grid
]
\addlegendimage{no markers, customG}
\addplot [thick, customG]
table [row sep=\\]{%
1112896216	0.0601851851851852 \\
1115488216	0.0738636363636364 \\
1118080216	0.0644230769230769 \\
1120672216	0.0528985507246377 \\
1123264216	0.0528985507246377 \\
1125856216	0.0528985507246377 \\
1128448216	0.0528985507246377 \\
1131040216	0.0528985507246377 \\
1133632216	0.0528985507246377 \\
1136224216	0.0540740740740741 \\
1138816216	0.0553030303030303 \\
1141408216	0.0711053652230123 \\
1144000216	0.0920716112531969 \\
1146592216	0.0951751487111699 \\
1149184216	0.132478632478632 \\
1151776216	0.302083333333333 \\
1154368216	0.279279279279279 \\
1156960216	0.255244755244755 \\
1159552216	0.208144796380091 \\
1162144216	0.201530612244898 \\
1164736216	0.175115207373272 \\
1167328216	0.165178571428571 \\
1169920216	0.158008658008658 \\
1172512216	0.152205882352941 \\
1175104216	0.124210526315789 \\
1177696216	0.120592743995912 \\
1180288216	0.128301886792453 \\
1182880216	0.130102040816327 \\
1185472216	0.119247699079632 \\
1188064216	0.110952040085898 \\
1190656216	0.120769741207697 \\
1193248216	0.113795295613477 \\
1195840216	0.108843537414966 \\
1198432216	0.112700228832952 \\
1201024216	0.102619497704564 \\
1203616216	0.107023411371237 \\
1206208216	0.125465838509317 \\
1208800216	0.105723905723906 \\
1211392216	0.10651289009498 \\
1213984216	0.0912395492548164 \\
1216576216	0.102189781021898 \\
1219168216	0.107313738892686 \\
1221760216	0.11950146627566 \\
1224352216	0.119174942704354 \\
1226944216	0.13893653516295 \\
1229536216	0.158514492753623 \\
1232128216	0.170829170829171 \\
1234720216	0.206119162640902 \\
1237312216	0.205115089514066 \\
1239904216	0.203014184397163 \\
1242496216	0.191326530612245 \\
1245088216	0.173831775700935 \\
1247680216	0.0847826086956522 \\
1250272216	0.098661028893587 \\
1252864216	0.1 \\
1255456216	0.100156494522692 \\
1258048216	0.0952380952380952 \\
1260640216	0.0958111548252719 \\
1263232216	0.0915032679738562 \\
1265824216	0.0979423868312757 \\
1268416216	0.0977225672877847 \\
1271008216	0.106095679012346 \\
1273600216	0.09383608815427 \\
1276192216	0.0933059407635679 \\
1278784216	0.087797619047619 \\
1281376216	0.0830792682926829 \\
1283968216	0.0831189392440135 \\
1286560216	0.0734848484848485 \\
1289152216	0.0986070381231672 \\
1291744216	0.101814516129032 \\
1294336216	0.104504504504505 \\
1296928216	0.107330286101236 \\
1299520216	0.105846774193548 \\
1302112216	0.0995989304812834 \\
1304704216	0.102238674422704 \\
1307296216	0.106141630205801 \\
1309888216	0.106731352334748 \\
1312480216	0.106731352334748 \\
1315072216	0.0997909407665505 \\
1317664216	0.102181562280084 \\
1320256216	0.0870748299319728 \\
1322848216	0.0845334269991804 \\
1325440216	0.0788272049987984 \\
1328032216	0.0681818181818182 \\
1330624216	0.0711500974658869 \\
1333216216	0.0723684210526316 \\
1335808216	0.064 \\
1338400216	0.0698668256807792 \\
1340992216	0.0622185460895138 \\
1343584216	0.0664084117321527 \\
1346176216	0.0668896321070234 \\
1348768216	0.0781558185404339 \\
1351360216	0.0808518005540166 \\
1353952216	0.0898713826366559 \\
1356544216	0.0895833333333333 \\
1359136216	0.0890127388535032 \\
1361728216	0.0923732094463802 \\
1364320216	0.0901907149912621 \\
1366912216	0.0924420355314664 \\
1369504216	0.0926829268292683 \\
1372096216	0.094492525570417 \\
1374688216	0.0959577433790624 \\
1377280216	0.0985926093173779 \\
1379872216	0.0861089256053851 \\
1382464216	0.0950649350649351 \\
1385056216	0.103597449908925 \\
1387648216	0.102573990166779 \\
1390240216	0.102788844621514 \\
1392832216	0.0968627450980392 \\
1395424216	0.10628594090819 \\
1398016216	0.0989661748469337 \\
1400608216	0.111121922740099 \\
1403200216	0.118258287976249 \\
1405792216	0.126414691437113 \\
1408384216	0.139428692789636 \\
1410976216	0.131932333449599 \\
1413568216	0.127745022643767 \\
1416160216	0.133938614816005 \\
1418752216	0.128160200250313 \\
1421344216	0.137565681798605 \\
1423936216	0.13033286451008 \\
1426528216	0.137495354886659 \\
1429120216	0.13441975308642 \\
1431712216	0.122408293460925 \\
1434304216	0.141505553270259 \\
1436896216	0.13455772113943 \\
1439488216	0.134370314842579 \\
1442080216	0.142067124332571 \\
1444672216	0.146236559139785 \\
1447264216	0.141940657578188 \\
1449856216	0.144639674051439 \\
1452448216	0.149659863945578 \\
1455040216	0.15590354767184 \\
1457632216	0.144983818770227 \\
1460224216	0.135501938819474 \\
1462816216	0.139196206818695 \\
1465408216	0.140102098695406 \\
1468000216	0.153 \\
1470592216	0.156683168316832 \\
1473184216	0.149425287356322 \\
1475776216	0.139765662362506 \\
1478368216	0.139421991420185 \\
1480960216	0.138748455229749 \\
1483552216	0.132905890023908 \\
1486144216	0.12979797979798 \\
1488736216	0.140559857057772 \\
1491328216	0.149031007751938 \\
1493920216	0.142072752230611 \\
1496512216	0.147047832585949 \\
1499104216	0.14170363797693 \\
1501696216	0.141763848396501 \\
1504288216	0.156431924882629 \\
1506880216	0.175959595959596 \\
1509472216	0.172475247524752 \\
1512064216	0.177147000932353 \\
1514656216	0.182397959183673 \\
1517248216	0.201684210526316 \\
1519840216	0.185781786941581 \\
1522432216	0.175669740887132 \\
1525024216	0.187170844744049 \\
1527616216	0.182199546485261 \\
1530208216	0.174418604651163 \\
1532800216	0.176928147659855 \\
1535392216	0.181377172887751 \\
1537984216	0.193370165745856 \\
1540576216	0.179298710688035 \\
1543168216	0.199893617021277 \\
};
\end{axis}
	\end{tikzpicture}
	\caption{Time series of different (network-analytic) measures for the time-stamped co-editing networks of an Open Source (left) and commercial software project (right). Results were generated using a rolling window analysis with a window size of 365 days and 30 day increments.}
	\label{fig:time_series_analysis}
\end{figure*}

A major advantage of \texttt{git2net} is its support for the extraction of \emph{dynamic} co-editing networks with high temporal resolution.
To showcase the benefits of such a temporal analysis for the two projects mentioned above, we have used \texttt{git2net}'s \texttt{python} API to extract a time-stamped co-editing network from the repositories of the two projects mentioned above.
We then used the temporal network analysis package \texttt{pathpy}~\cite{pathpy} to apply a rolling window analysis, which provided us with a time series of network analytic measures.
Figure \ref{fig:time_series_analysis} shows the resulting time series for four measures both for the Open Source project \texttt{igraph} as well as the commercial software project.
The first row gives the number of developers working on the projects in a 365-day sliding window. 
The number of unique co-editing relations between these developers, shown in the second row, can be used to proxy the amount of collaboration on joint code regions taking place in a project in a given time window.
We observe that the number of such collaborations relative to the number of developers is considerably higher for the commercial software project compared to the Open Source project. 
This finding is further corroborated by the mean out-degree of nodes shown in the third row.
This suggests that on average developers in \texttt{igraph} edit the code of one to two other developers, while for the commercial software project each developer has to coordinate his or her changes with four to eight other team members.
It is a remarkable finding for the commercial software project that both the number of unique directed edges and the mean out-degree  decline from 2013 onwards, despite the growing number of developers.
This could mark a change in the software development processes and/or the social organisation of teams.
While a first feedback from the project managers suggests that this could be related to a change in the adoption of an agile development model, testing this hypothesis requires a separate in-depth study.
Finally, in the fourth row in Figure \ref{fig:time_series_analysis} we report the evolution of normalised (total) degree centralisation over time~\cite{Freeman1978}.
A minimum value of zero indicates that all nodes in the network have the same degree, while a maximum value of one corresponds to a perfect star network where all nodes except a hub node have degree one.
We find that \texttt{igraph} exhibits considerably larger degree centralisation than the commercial software project, which is likely related to previous findings of highly skewed distributions of code contributions in Open Source projects~\cite{Scholtes2016, mockus2002two,Lin2015}.

\subsection{Editing of Own vs. Foreign Code}\label{sec:results:ownership}

In a final experiment, we showcase how \texttt{git2net} can be used to analyse temporal co-editing patterns in software development teams.
To this end, we extend our analysis of the mere \emph{topological} dimension of co-editing relations performed in previous sections, to use additional information on the Levenshtein distance associated with these relations.
The Levenshtein distance between two source code versions captures the number of characters one has to type to transform one string into another string.
It has been used as a proxy for development effort associated with commits~\cite{Scholtes2016}.
Extending this approach, an interesting aspect of our methodology is that it allows us to distinguish between (i)  the cumulative Levenshtein distance of code edits made in a developer's \emph{own} code and (ii) the cumulative Levenshtein distance of edits made in \emph{foreign} code, i.e. code originally written by other developers.
This enables us to calculate, for each time window in the commit history of a project, the relative proportion of development effort falling into these two categories.

\begin{figure*}[t!]
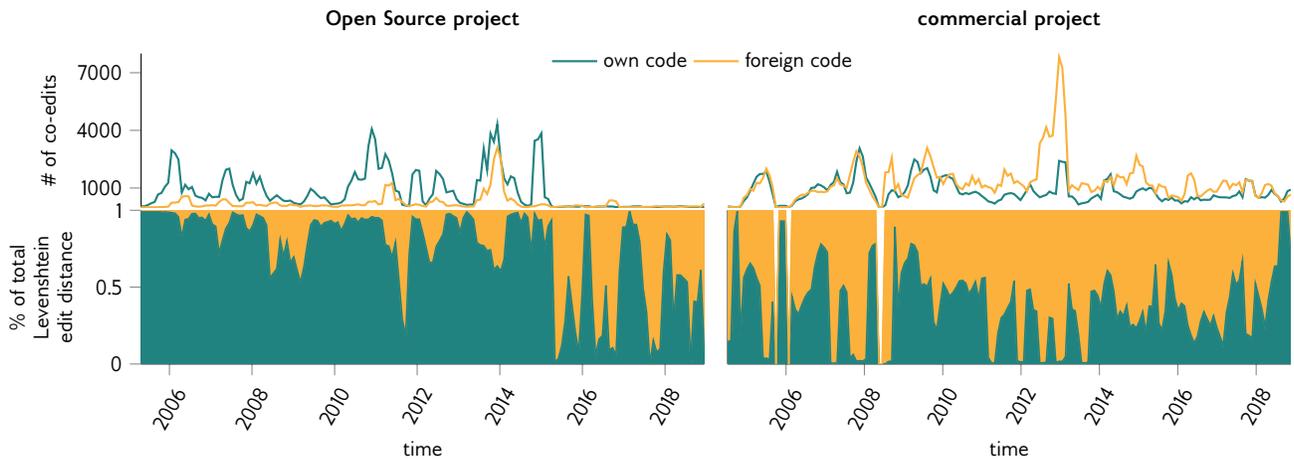

    \centering
    \begin{tikzpicture}\footnotesize\sffamily
    	\begin{axis}[
at={(0,0)},
width=.5\textwidth,
height=.2\textwidth,
ylabel={\# of co-edits},
xmin=1114430052, xmax=1544702052,
ymin=0, ymax=8000,
tick align=outside,
tick pos=left,
x grid style={white!69.01960784313725!black},
y grid style={white!69.01960784313725!black},
axis y line*=left,
x axis line style={draw opacity=0},
xtick = \empty,
legend entries={{own code},{foreign code}},
legend cell align={left},
legend style={at={(1,.95)},anchor=center,legend columns=3,legend cell align=left,align=left,fill=none,draw=none},
ytick = {1000, 4000, 7000},
yticklabels = {1000, 4000, 7000},
title={\textbf{Open Source project}}
]
\addplot [thick, customG]
table [row sep=\\]{%
1114430052	47 \\
1117022052	50 \\
1119614052	120 \\
1122206052	208 \\
1124798052	276 \\
1127390052	667 \\
1129982052	743 \\
1132574052	1058 \\
1135166052	1273 \\
1137758052	2961 \\
1140350052	2808 \\
1142942052	2481 \\
1145534052	772 \\
1148126052	1170 \\
1150718052	935 \\
1153310052	1056 \\
1155902052	603 \\
1158494052	523 \\
1161086052	486 \\
1163678052	356 \\
1166270052	700 \\
1168862052	517 \\
1171454052	543 \\
1174046052	539 \\
1176638052	1412 \\
1179230052	1953 \\
1181822052	2009 \\
1184414052	1157 \\
1187006052	631 \\
1189598052	344 \\
1192190052	441 \\
1194782052	1320 \\
1197374052	1247 \\
1199966052	1761 \\
1202558052	1229 \\
1205150052	1585 \\
1207742052	1022 \\
1210334052	587 \\
1212926052	335 \\
1215518052	485 \\
1218110052	622 \\
1220702052	478 \\
1223294052	336 \\
1225886052	321 \\
1228478052	345 \\
1231070052	229 \\
1233662052	180 \\
1236254052	146 \\
1238846052	304 \\
1241438052	584 \\
1244030052	958 \\
1246622052	821 \\
1249214052	626 \\
1251806052	475 \\
1254398052	521 \\
1256990052	334 \\
1259582052	141 \\
1262174052	177 \\
1264766052	193 \\
1267358052	250 \\
1269950052	412 \\
1272542052	1077 \\
1275134052	1386 \\
1277726052	1828 \\
1280318052	1452 \\
1282910052	1430 \\
1285502052	1489 \\
1288094052	3170 \\
1290686052	4067 \\
1293278052	3536 \\
1295870052	2000 \\
1298462052	2050 \\
1301054052	2758 \\
1303646052	2429 \\
1306238052	1760 \\
1308830052	861 \\
1311422052	801 \\
1314014052	132 \\
1316606052	87 \\
1319198052	62 \\
1321790052	1715 \\
1324382052	1938 \\
1326974052	1899 \\
1329566052	307 \\
1332158052	103 \\
1334750052	447 \\
1337342052	676 \\
1339934052	1891 \\
1342526052	1727 \\
1345118052	1468 \\
1347710052	775 \\
1350302052	817 \\
1352894052	859 \\
1355486052	424 \\
1358078052	227 \\
1360670052	178 \\
1363262052	123 \\
1365854052	140 \\
1368446052	130 \\
1371038052	1413 \\
1373630052	1382 \\
1376222052	3043 \\
1378814052	1946 \\
1381406052	3830 \\
1383998052	3411 \\
1386590052	4315 \\
1389182052	2481 \\
1391774052	1502 \\
1394366052	1190 \\
1396958052	1609 \\
1399550052	1514 \\
1402142052	766 \\
1404734052	284 \\
1407326052	139 \\
1409918052	69 \\
1412510052	142 \\
1415102052	3458 \\
1417694052	3534 \\
1420286052	3842 \\
1422878052	518 \\
1425470052	428 \\
1428062052	46 \\
1430654052	1 \\
1433246052	5 \\
1435838052	17 \\
1438430052	36 \\
1441022052	43 \\
1443614052	35 \\
1446206052	37 \\
1448798052	26 \\
1451390052	69 \\
1453982052	48 \\
1456574052	50 \\
1459166052	4 \\
1461758052	21 \\
1464350052	34 \\
1466942052	35 \\
1469534052	49 \\
1472126052	37 \\
1474718052	39 \\
1477310052	8 \\
1479902052	4 \\
1482494052	17 \\
1485086052	17 \\
1487678052	17 \\
1490270052	42 \\
1492862052	42 \\
1495454052	50 \\
1498046052	8 \\
1500638052	11 \\
1503230052	3 \\
1505822052	15 \\
1508414052	12 \\
1511006052	12 \\
1513598052	32 \\
1516190052	33 \\
1518782052	33 \\
1521374052	24 \\
1523966052	40 \\
1526558052	65 \\
1529150052	42 \\
1531742052	25 \\
1534334052	4 \\
1536926052	14 \\
1539518052	14 \\
1542110052	10 \\
1544702052	19 \\
};
\addplot [thick, customY]
table [row sep=\\]{%
1114430052	0 \\
1117022052	0 \\
1119614052	0 \\
1122206052	0 \\
1124798052	0 \\
1127390052	0 \\
1129982052	9 \\
1132574052	15 \\
1135166052	33 \\
1137758052	236 \\
1140350052	243 \\
1142942052	265 \\
1145534052	540 \\
1148126052	577 \\
1150718052	567 \\
1153310052	85 \\
1155902052	44 \\
1158494052	29 \\
1161086052	46 \\
1163678052	88 \\
1166270052	95 \\
1168862052	106 \\
1171454052	81 \\
1174046052	271 \\
1176638052	412 \\
1179230052	409 \\
1181822052	226 \\
1184414052	63 \\
1187006052	59 \\
1189598052	46 \\
1192190052	58 \\
1194782052	236 \\
1197374052	232 \\
1199966052	264 \\
1202558052	102 \\
1205150052	129 \\
1207742052	91 \\
1210334052	71 \\
1212926052	190 \\
1215518052	236 \\
1218110052	243 \\
1220702052	87 \\
1223294052	69 \\
1225886052	69 \\
1228478052	153 \\
1231070052	108 \\
1233662052	96 \\
1236254052	21 \\
1238846052	74 \\
1241438052	79 \\
1244030052	93 \\
1246622052	57 \\
1249214052	63 \\
1251806052	89 \\
1254398052	95 \\
1256990052	83 \\
1259582052	39 \\
1262174052	22 \\
1264766052	20 \\
1267358052	86 \\
1269950052	92 \\
1272542052	98 \\
1275134052	99 \\
1277726052	122 \\
1280318052	150 \\
1282910052	101 \\
1285502052	107 \\
1288094052	289 \\
1290686052	286 \\
1293278052	321 \\
1295870052	144 \\
1298462052	182 \\
1301054052	1156 \\
1303646052	1128 \\
1306238052	1266 \\
1308830052	369 \\
1311422052	436 \\
1314014052	267 \\
1316606052	118 \\
1319198052	62 \\
1321790052	106 \\
1324382052	154 \\
1326974052	126 \\
1329566052	71 \\
1332158052	36 \\
1334750052	269 \\
1337342052	292 \\
1339934052	491 \\
1342526052	300 \\
1345118052	271 \\
1347710052	81 \\
1350302052	70 \\
1352894052	98 \\
1355486052	105 \\
1358078052	74 \\
1360670052	41 \\
1363262052	25 \\
1365854052	38 \\
1368446052	45 \\
1371038052	369 \\
1373630052	394 \\
1376222052	963 \\
1378814052	685 \\
1381406052	1180 \\
1383998052	2494 \\
1386590052	3085 \\
1389182052	2551 \\
1391774052	804 \\
1394366052	343 \\
1396958052	458 \\
1399550052	340 \\
1402142052	150 \\
1404734052	36 \\
1407326052	9 \\
1409918052	9 \\
1412510052	46 \\
1415102052	51 \\
1417694052	104 \\
1420286052	155 \\
1422878052	149 \\
1425470052	97 \\
1428062052	8 \\
1430654052	6 \\
1433246052	31 \\
1435838052	32 \\
1438430052	42 \\
1441022052	21 \\
1443614052	44 \\
1446206052	98 \\
1448798052	90 \\
1451390052	70 \\
1453982052	5 \\
1456574052	10 \\
1459166052	10 \\
1461758052	28 \\
1464350052	39 \\
1466942052	35 \\
1469534052	66 \\
1472126052	348 \\
1474718052	356 \\
1477310052	306 \\
1479902052	9 \\
1482494052	1 \\
1485086052	1 \\
1487678052	0 \\
1490270052	4 \\
1492862052	4 \\
1495454052	16 \\
1498046052	15 \\
1500638052	19 \\
1503230052	20 \\
1505822052	31 \\
1508414052	54 \\
1511006052	41 \\
1513598052	32 \\
1516190052	8 \\
1518782052	22 \\
1521374052	28 \\
1523966052	29 \\
1526558052	37 \\
1529150052	33 \\
1531742052	29 \\
1534334052	44 \\
1536926052	41 \\
1539518052	45 \\
1542110052	23 \\
1544702052	201 \\
};
,\end{axis}
    	\input{figures/igraph_self_changes_dist_norm.tex}
    	\begin{axis}[
at={(.43\textwidth,0)},
width=.5\textwidth,
height=.2\textwidth,
ymin=0, ymax=8000,
xmin=1089054280, xmax=1542654280,
tick align=outside,
tick pos=left,
hide axis,
xtick = \empty,
ytick = \empty,
title={\textbf{commercial project}}
]
\addplot [thick, customG]
table [row sep=\\]{%
1089054280	31 \\
1091646280	28 \\
1094238280	4 \\
1096830280	3 \\
1099422280	40 \\
1102014280	326 \\
1104606280	514 \\
1107198280	859 \\
1109790280	1235 \\
1112382280	1589 \\
1114974280	1728 \\
1117566280	1733 \\
1120158280	1861 \\
1122750280	1340 \\
1125342280	670 \\
1127934280	0 \\
1130526280	72 \\
1133118280	72 \\
1135710280	72 \\
1138302280	0 \\
1140894280	89 \\
1143486280	316 \\
1146078280	410 \\
1148670280	455 \\
1151262280	553 \\
1153854280	692 \\
1156446280	1009 \\
1159038280	1019 \\
1161630280	1196 \\
1164222280	1138 \\
1166814280	1030 \\
1169406280	875 \\
1171998280	1160 \\
1174590280	1311 \\
1177182280	1846 \\
1179774280	1302 \\
1182366280	1186 \\
1184958280	729 \\
1187550280	757 \\
1190142280	1273 \\
1192734280	2384 \\
1195326280	3062 \\
1197918280	2638 \\
1200510280	1557 \\
1203102280	1250 \\
1205694280	896 \\
1208286280	533 \\
1210878280	0 \\
1213470280	0 \\
1216062280	106 \\
1218654280	680 \\
1221246280	885 \\
1223838280	838 \\
1226430280	523 \\
1229022280	604 \\
1231614280	781 \\
1234206280	1134 \\
1236798280	1923 \\
1239390280	2487 \\
1241982280	2356 \\
1244574280	1775 \\
1247166280	1954 \\
1249758280	2035 \\
1252350280	1599 \\
1254942280	883 \\
1257534280	774 \\
1260126280	1539 \\
1262718280	1624 \\
1265310280	1674 \\
1267902280	1559 \\
1270494280	1477 \\
1273086280	1092 \\
1275678280	727 \\
1278270280	668 \\
1280862280	739 \\
1283454280	684 \\
1286046280	760 \\
1288638280	807 \\
1291230280	724 \\
1293822280	583 \\
1296414280	455 \\
1299006280	303 \\
1301598280	263 \\
1304190280	188 \\
1306782280	332 \\
1309374280	399 \\
1311966280	650 \\
1314558280	706 \\
1317150280	735 \\
1319742280	689 \\
1322334280	533 \\
1324926280	372 \\
1327518280	623 \\
1330110280	899 \\
1332702280	1194 \\
1335294280	849 \\
1337886280	665 \\
1340478280	516 \\
1343070280	585 \\
1345662280	639 \\
1348254280	817 \\
1350846280	730 \\
1353438280	694 \\
1356030280	2412 \\
1358622280	2362 \\
1361214280	2343 \\
1363806280	562 \\
1366398280	564 \\
1368990280	404 \\
1371582280	145 \\
1374174280	202 \\
1376766280	233 \\
1379358280	277 \\
1381950280	455 \\
1384542280	556 \\
1387134280	651 \\
1389726280	439 \\
1392318280	1398 \\
1394910280	1474 \\
1397502280	1713 \\
1400094280	830 \\
1402686280	784 \\
1405278280	671 \\
1407870280	544 \\
1410462280	499 \\
1413054280	442 \\
1415646280	518 \\
1418238280	817 \\
1420830280	914 \\
1423422280	983 \\
1426014280	889 \\
1428606280	1006 \\
1431198280	980 \\
1433790280	695 \\
1436382280	452 \\
1438974280	312 \\
1441566280	439 \\
1444158280	380 \\
1446750280	498 \\
1449342280	346 \\
1451934280	354 \\
1454526280	201 \\
1457118280	311 \\
1459710280	291 \\
1462302280	520 \\
1464894280	410 \\
1467486280	538 \\
1470078280	357 \\
1472670280	394 \\
1475262280	381 \\
1477854280	397 \\
1480446280	622 \\
1483038280	556 \\
1485630280	519 \\
1488222280	514 \\
1490814280	512 \\
1493406280	484 \\
1495998280	579 \\
1498590280	623 \\
1501182280	822 \\
1503774280	525 \\
1506366280	1458 \\
1508958280	1382 \\
1511550280	1364 \\
1514142280	618 \\
1516734280	494 \\
1519326280	523 \\
1521918280	501 \\
1524510280	641 \\
1527102280	903 \\
1529694280	691 \\
1532286280	574 \\
1534878280	296 \\
1537470280	463 \\
1540062280	840 \\
1542654280	913 \\
};
\addplot [thick, customY]
table [row sep=\\]{%
1089054280	22 \\
1091646280	17 \\
1094238280	3 \\
1096830280	0 \\
1099422280	34 \\
1102014280	494 \\
1104606280	626 \\
1107198280	965 \\
1109790280	864 \\
1112382280	1274 \\
1114974280	1245 \\
1117566280	1524 \\
1120158280	2004 \\
1122750280	1660 \\
1125342280	1022 \\
1127934280	0 \\
1130526280	4 \\
1133118280	4 \\
1135710280	4 \\
1138302280	0 \\
1140894280	125 \\
1143486280	402 \\
1146078280	540 \\
1148670280	652 \\
1151262280	639 \\
1153854280	758 \\
1156446280	854 \\
1159038280	858 \\
1161630280	792 \\
1164222280	802 \\
1166814280	789 \\
1169406280	825 \\
1171998280	1064 \\
1174590280	1134 \\
1177182280	1509 \\
1179774280	1186 \\
1182366280	1120 \\
1184958280	1021 \\
1187550280	1694 \\
1190142280	2360 \\
1192734280	2916 \\
1195326280	2650 \\
1197918280	2085 \\
1200510280	1281 \\
1203102280	923 \\
1205694280	563 \\
1208286280	308 \\
1210878280	0 \\
1213470280	0 \\
1216062280	1812 \\
1218654280	1827 \\
1221246280	2565 \\
1223838280	909 \\
1226430280	1201 \\
1229022280	631 \\
1231614280	580 \\
1234206280	496 \\
1236798280	1009 \\
1239390280	1512 \\
1241982280	1671 \\
1244574280	1798 \\
1247166280	2466 \\
1249758280	3079 \\
1252350280	2652 \\
1254942280	2080 \\
1257534280	1540 \\
1260126280	1576 \\
1262718280	1206 \\
1265310280	1526 \\
1267902280	1721 \\
1270494280	1740 \\
1273086280	1213 \\
1275678280	1015 \\
1278270280	937 \\
1280862280	1241 \\
1283454280	1129 \\
1286046280	1618 \\
1288638280	1219 \\
1291230280	1460 \\
1293822280	1029 \\
1296414280	1134 \\
1299006280	1048 \\
1301598280	996 \\
1304190280	757 \\
1306782280	1178 \\
1309374280	1115 \\
1311966280	1960 \\
1314558280	1637 \\
1317150280	1752 \\
1319742280	1257 \\
1322334280	1362 \\
1324926280	1225 \\
1327518280	1096 \\
1330110280	832 \\
1332702280	1295 \\
1335294280	1395 \\
1337886280	1445 \\
1340478280	3331 \\
1343070280	3618 \\
1345662280	4167 \\
1348254280	3673 \\
1350846280	3729 \\
1353438280	5621 \\
1356030280	7812 \\
1358622280	7260 \\
1361214280	4732 \\
1363806280	877 \\
1366398280	1149 \\
1368990280	1014 \\
1371582280	898 \\
1374174280	1249 \\
1376766280	1220 \\
1379358280	1177 \\
1381950280	795 \\
1384542280	858 \\
1387134280	1064 \\
1389726280	902 \\
1392318280	1380 \\
1394910280	1307 \\
1397502280	1559 \\
1400094280	1126 \\
1402686280	1688 \\
1405278280	1552 \\
1407870280	1578 \\
1410462280	1201 \\
1413054280	1754 \\
1415646280	1788 \\
1418238280	2675 \\
1420830280	2203 \\
1423422280	2123 \\
1426014280	1299 \\
1428606280	1138 \\
1431198280	1121 \\
1433790280	724 \\
1436382280	1184 \\
1438974280	1076 \\
1441566280	1045 \\
1444158280	486 \\
1446750280	614 \\
1449342280	571 \\
1451934280	530 \\
1454526280	331 \\
1457118280	691 \\
1459710280	1650 \\
1462302280	1764 \\
1464894280	1427 \\
1467486280	587 \\
1470078280	562 \\
1472670280	717 \\
1475262280	857 \\
1477854280	949 \\
1480446280	804 \\
1483038280	816 \\
1485630280	1107 \\
1488222280	1382 \\
1490814280	1368 \\
1493406280	966 \\
1495998280	1217 \\
1498590280	1150 \\
1501182280	1294 \\
1503774280	796 \\
1506366280	1421 \\
1508958280	1418 \\
1511550280	1335 \\
1514142280	586 \\
1516734280	487 \\
1519326280	739 \\
1521918280	1157 \\
1524510280	1206 \\
1527102280	1126 \\
1529694280	724 \\
1532286280	569 \\
1534878280	356 \\
1537470280	300 \\
1540062280	565 \\
1542654280	648 \\
};
\end{axis}
    	\input{figures/genua_self_changes_dist_norm.tex}
    \end{tikzpicture}
    \caption{Editing of own and foreign code for Open Source and commercial project over time. The total number of edited blocks is shown above whereas the bottom figures show proportions of the total Levenshtein edit distance. Results are computed on a 90 day rolling window with 30 day increments.}\label{fig:time_series}
\end{figure*}

Figure \ref{fig:time_series} shows the result of this analysis for the two projects introduced above, where the top-part of the figure reports the total number of (unweighted) co-edit relations, while the bottom part shows the relative proportion of the total Levenshtein distance of own code changes vs. foreign code changes.
This analysis highlights considerable project- and time-dependent differences.
For the Open Source project \texttt{igraph}, during a first phase from 2006 to 2015, the majority of code edits take place in code previously written by the same developer.
This indicates a strict notion of code ``ownership'', where developers rarely touch code written by others. 
For the commercial software project we observe a completely different dynamics, where for the majority of time windows development effort is dominated by \emph{foreign} code edits.
We hypothesise that this finding is likely related to code changes triggered by the specific implementation of the code review process in the commercial software project~\cite{Beller2014}. 
While an in-depth study of this claim goes beyond the scope of this tool paper, this finding highlights a specific research question that can be addressed with our tool in future work.
%
\section{Conclusion and Outlook}
\label{sec:conclusion}

Over the past two decades, the analysis of co-authorship, co-commit, or co-editing networks in software development teams has experienced huge interest from the empirical software engineering and repository mining community.
Exemplary studies have shown that the analysis of such collaboration networks helps to assess the time-evolving social structure of teams~\cite{Scholtes2016, madey2002}, predict software defects~\cite{Meneely2008}, categorise developer roles~\cite{pohl2008dynamic}, identify communities~\cite{Joblin2015}, or study knowledge spillover across individuals, teams, and projects~\cite{vonKrogh2006, vijayaraghavan2015quantifying, Huang2005, Cohen2018}.
Most of these studies have employed definitions of co-authorship networks which assume that developers are linked if they edited a common file, module, or binary.
However, such coarse-grained definitions have been shown to neglect information on the microscopic patterns of collaborations contained in the time-ordered sequence of lines of code edited by developers~\cite{Joblin2015,Scholtes2016}.

To facilitate data-driven studies of developer networks that take advantage of this detailed information, we have introduced \texttt{git2net}, a \texttt{python} package for the mining of fine-grained and time-stamped collaboration networks from large \texttt{git} repositories.
Going beyond previous works, we adopt text mining techniques to assess (a) the development effort of an edit in terms of the Levenshtein distance between the version before and after the commit, and (b) the entropy of file modifications, which can be used to filter out changes in text-encoded binary data.
Thanks to a parallel processing model our tool exhibits a linear speed up for an increasing number of processing cores.
This makes \texttt{git2net} suitable to analyse \texttt{git} repositories with hundreds of thousands of commits and millions of lines of code.

Apart from a description of our tool, we have reported results of a case study using the repositories of an Open Source and a commercial software project.
While the results are rather anecdotal and should thus not be generalised to other projects, this case study is meant to demonstrate that the presented tool simplifies the construction and analysis of dynamic developer collaboration networks and co-editing behaviour.
It further showcases scenarios where our tool can be useful and highlights interesting research questions that we will address in future works.

Extending the analysis presented in \cite{Joblin2015}, in section \ref{sec:results:coauthorship} we report on a small comparative study of a file- vs. line-based construction of co-editing networks.
A future systematic study of the differences between these approaches would be important.
This should highlight in which case we need fine-grained methods and in which other cases coarse-grained notions of collaboration may be sufficient.
Given the large number of studies using coarse-grained definitions of collaboration networks, such a study could make a substantial methodological contribution to the repository mining literature.

The results presented in section \ref{sec:results:temporal} indicate topological differences between co-editing networks that are potentially linked to (a) the difference between Open Source and commercial software projects, and (b) the adoption of an agile development process in the commercial software project. 
These hypotheses must be tested in a larger corpus of projects that differ in these two dimensions.
To support such a study, we recently mined co-editing relationships from the full \texttt{git} commit history of Linux\footnote{\url{https://github.com/torvalds/linux}}, comprising more than 800,000 commits over a period of 18 years.
Running \texttt{git2net} on a machine with 16 processing cores, we were able to complete the extraction of more than 60 million time-stamped co-editing relations in four days.

In section \ref{sec:results:ownership} we further demonstrate that the information extracted by our tool can be used to generate a time-resolved breakdown of developer effort into (a) the revision of code authored by the developer him or herself vs. (b) the revision of code written by other team members. 
We currently work on a more systematic analysis of this interesting aspect of collaboration in development teams.
Specifically, we study how collaboration is related to team size, project types, release schedules, code review processes, or the difference between Open Source and industrial projects.

Finally, a key advantage of our tool is that it provides a simple method to extract fine-grained collaboration networks at high temporal resolution from any \texttt{git} repository.
Publicly available repositories cover a variety of different collaborative tasks, like software development, manuscript editing, web content management, etc. \cite{kalliamvakou2016depth}.
Our tool efficiently utilises the large number of such repositories and thus opens up a massive new source of high-resolution data on human collaboration patterns.

The fact that the resulting dynamic collaboration networks can be cross-referenced with project-related information (project success, organisational structures and project culture, developer roles, etc.) is likely to be of value for researchers in computational social science and organisational theory.
We further expect the resulting corpus of data to be of considerable interest for the network science and social network analysis community, which have recently moved beyond moving window analyses, developing techniques that incorporate the chronological ordering of interactions in high-resolution time series data~\cite{newman2018networks,Holme2015,Scholtes2017}.
We thus hope that the tool and analyses presented in our work will serve the growing community of interdisciplinary researchers working at the intersection of data science, (social) network analysis, computational social science and empirical software engineering.

\section*{Acknowledgment}
Ingo Scholtes acknowledges financial support by the Swiss National Science Foundation through grant 176938.
We thank Alexander von Gernler as well as all other members of the software company \textsc{Genua} for allowing us to validate our tool in a large commercial software project.
All authors express their thanks to the anonymous reviewers of the manuscript.

\section*{Tool availability, archival, and reproducibility}

The tool presented in this work is available as Open Source software package on \texttt{gitHub}\footnote{\url{https://github.com/gotec/git2net}}.
\texttt{git2net} is further available via the \texttt{python} package index \texttt{pypi}, enabling users to simply install and update it via the package management tool \texttt{pip}.
To support the reproducibility of our work, we have permanently archived the version of our tool that was used to obtain the results presented in this paper on the open-access repository \texttt{zenodo.org}~\cite{Gote2019}.

\texttt{git2net} comes with unit tests and a comprehensive in-line documentation.
To support users in developing their first analysis, we further provide access to interactive \texttt{jupyter} notebooks, which allow to reproduce our approach.

Since the submission of this paper, the following additional features have been added to the release version of \texttt{git2net}:
\begin{itemize}
	\item Extraction of \emph{line-editing networks}, where nodes represent states of content lines of files, while edges link consecutive versions.
	\item Detection of copying and moving lines both within and between files for the file-based approach via the \texttt{git blame -C} option.
	\item Extraction of edits rather than co-edits in the \texttt{sqlite} database. With this also pure additions are listed in the database allowing users of \texttt{git2net} to implement own co-editing measures, e.g. based on the distance (in line numbers) of an addition to other lines. 
\end{itemize}


\footnotesize

\bibliographystyle{IEEEtran}
\bibliography{abstracts}

\end{document}